\documentclass{aa_nolineno}  

\usepackage{longtable,array}
\usepackage{graphicx}
\usepackage{txfonts}
\usepackage[utf8]{inputenc} 
\usepackage{txfonts}
\usepackage{CJKutf8}
\usepackage{upgreek}
\usepackage{amsmath}
\usepackage{bm}
\usepackage{subcaption}
\usepackage{orcidlink}

\usepackage{hyperref}

\def\txt{\mathrm}
\def\microm{\mathrm {\upmu m}}
\def\microbar{\mathrm{\upmu bar}}
\def\h{\mathrm{h}}
\def\m{\mathrm{m}}
\def\s{\mathrm{s}}
\def\cm{\mathrm{cm}}
\def\km{\mathrm{km}}
\def\J{\mathrm{J}}
\def\K{\mathrm{K}}

\newcommand{\gkai}[1]{\begin{CJK}{UTF8}{gkai}{#1}\end{CJK}}

\newcommand{\bkai}[1]{\begin{CJK}{UTF8}{bkai}{#1}\end{CJK}}

\graphicspath{{./}{figures/}}

\begin{document}

   \title{Reconciling results of 2019 and 2020 stellar occultations on Pluto's atmosphere}

   \subtitle{New constraints from both the 5 September 2019 event and consistency analysis}

   \author{
      {Ye Yuan} {(\gkai{袁烨})}
      \inst{1} \orcidlink{0000-0002-4686-8548}
      \and
      {Fan Li} {(\gkai{李凡})}
      \inst{1} \orcidlink{0000-0002-2997-3098}
      \and 
      {Yanning Fu} {(\gkai{傅燕宁})}
      \inst{1} 
      \and 
      {Jian Chen} {(\gkai{陈健})}
      \inst{1,2} \orcidlink{0009-0006-9369-0806}
      \and
      {Wei Tan} {(\gkai{谭巍})}
      \inst{3}
      \and 
      {Shuai Zhang} {(\gkai{张帅})}
      \inst{4} \orcidlink{0000-0002-4799-0780}
      \and 
      {Wei Zhang} {(\gkai{张伟})}
      \inst{1}
      \and 
      {Chen Zhang} {(\gkai{张晨})}
      \inst{1,2} \orcidlink{0000-0002-9583-263X}
      \and 
      {Qiang Zhang} {(\gkai{张强})}
      \inst{4}
      \and 
      {Jiahui Ye} {(\gkai{叶嘉晖})}
      \inst{5}
      \and 
      {Delai Li} {(\gkai{李德铼})}
      \inst{5}
      \and 
      {Yijing Zhu} {(\gkai{朱一静})}
      \inst{6}
      \and 
      {Zhensen Fu} {(\gkai{傅震森})}
      \inst{7,4}
      \and  
      {Ansheng Zhu} {(\gkai{朱安生})}
      \inst{1,2} \orcidlink{0009-0007-5505-526X}
      \and 
      {Yue Chen} {(\gkai{陈悦})}
      \inst{1,2} \orcidlink{0000-0002-6050-9920}
      \and 
      {Jun Xu} {(\gkai{许军})}
      \inst{8}
      \and
      {Yang Zhang} {(\gkai{张}\bkai{暘})}
      \inst{1}
   }

   \institute{
      {Purple Mountain Observatory, Chinese Academy of Sciences, No. 10 Yuanhua Road, Nanjing 210033, China}\\
      \email{yuanye@pmo.ac.cn} 
      \and
      {School of Astronomy and Space Science, University of Science and Technology of China, No. 96 Jinzhai Road, Hefei, Anhui 230026, China} 
      \and
      {Hunan Astronomical Association, Changsha, Hunan 410000, China} 
      \and
      {Department of Space Sciences and Astronomy, Hebei Normal University, No. 20 Road East. 2nd Ring South, Shijiazhuang, Hebei 050024, China} 
      \and
      {Shenzhen Astronomical Observatory, Tianwen Road, Shenzhen, Guangdong 518040, China} 
      \and
      {Shenzhen Astronomical Society, 22c Seascape Square Taizi Road, Shenzhen, Guangdong 518040, China} 
      \and
      {Shanghai Astronomical Observatory, Chinese Academy of Sciences, No. 80 Nandan Road, Shanghai 200030, China} 
      \and
      {Nanjing Amateur Astronomers Association, Nanjing 210000, China} 
   }

   \date{}


   \abstract
   {
      A stellar occultation by Pluto on 5 September 2019 yielded positive detections at two separate stations.
      Using an approach consistent with comparable studies, we derived a surface pressure of $11.478 \pm 0.55~\microbar$ for Pluto's atmosphere from the observations of this event. 
      In addition, to avoid potential method inconsistencies when comparing with historical pressure measurements, we reanalyzed the data for the 15 August 2018 and {17 July} 2019 events. 
      All the new measurements provide a bridge between the two different perspectives on the pressure variation since 2015: a rapid pressure drop from previous studies of the 15 August 2018 and {17 July} 2019 events and a plateau phase from that of the 6 June 2020 event.
      The pressure measurement from the 5 September 2019 event aligns with those from 2016, 2018, and 2020, supporting the latter perspective.
      While the measurements from the 4 June 2011 and 17 July 2019 events suggest probable V-shaped pressure variations that are unaccounted for by the volatile transport model (VTM), the VTM remains applicable on average.
      Furthermore, the validity of the V-shaped variations is debatable given the stellar faintness of the 4 June 2011 event and the grazing single-chord geometry of the 17 July 2019 event.
      To reveal and understand all of the significant pressure variations of Pluto's atmosphere, it is essential to provide constraints on both the short-term and long-term evolution of the interacting atmosphere and surface by continuous pressure monitoring through occultation observations whenever possible, and to complement these with frequent spectroscopy and photometry of the surface.
   }


   \authorrunning{Ye Yuan et al.}
   \maketitle

   \section{Introduction}
   \label{sect:introduction}

   Pluto's atmosphere was discovered during the 1985 stellar occultation \citep{Brosch1995}, and since then, stellar occultations have played a crucial role in studying its structure, composition, and evolution over time \citep{Hubbard1988,Elliot1989,Elliot2003a,Yelle1997,Sicardy2003,Sicardy2011a,Sicardy2016,Sicardy2021,Pasachoff2005,Pasachoff2017,Young2008,Young2021,Rannou2009,Person2013,Person2021,Olkin2015,Bosh2015,Gulbis2015,DiasOliveira2015,Meza2019,Arimatsu2020}.
   A compilation of 12 occultations observed between 1988 and 2016 revealed a three-fold monotonic increase in the atmospheric pressure of Pluto during that period \citep{Meza2019}.
   This increase can be explained by the volatile transport model (VTM) of the Laboratoire de M\'et\'eorologie Dynamique (LMD) \citep{Bertrand2016,Forget2017,Bertrand2018,Bertrand2019}, which was subsequently  fine-tuned by \citet{Meza2019}.
   This model provides a framework for simulating the volatile cycles on Pluto over both seasonal and astronomical timescales, allowing us to explore the long-term evolution of Pluto's atmosphere and its response to seasonal variations over its 248 year heliocentric orbital period \citep{Meza2019}. 
   According to the LMD VTM in \citet{Meza2019} (VTM19, hereafter), Pluto's atmospheric pressure is expected to have reached its peak around the year 2020.
   The pressure increase is attributed to the progression of summer over the northern hemisphere of Pluto, exposing Sputnik Planitia (SP)\footnote{Sputnik Planitia is a large plain or basin on Pluto discovered by the New Horizons spacecraft \citep{Stern2015}.} to solar radiation. 
   The surface of SP, which is composed of nitrogen ($\txt{N_2}$), methane ($\txt{CH_4}$), and carbon monoxide ($\txt{CO}$) ices, is believed to sublimate and release volatile gases into the atmosphere during this period, leading to a pressure increase.
   After reaching its peak, the model predicts a gradual decline in pressure over the next two centuries under the combined effects of Pluto's recession from the Sun and the prevalence of the winter season over SP.  

   On one hand, the VTM19 remains consistent with the analysis of \citet{Sicardy2021}  of the 6 June 2020 occultation observed at Devasthal, where two colocated telescopes were used. 
   This latter analysis suggests that Pluto's atmosphere has been in a plateau phase since mid-2015, which aligns with the model predictions that the atmospheric pressure reached its peak around 2020.
    
   On the other hand, the \citet{Arimatsu2020} analysis of the 17 July 2019 occultation observed by a single telescope (TUHO) suggests a rapid pressure decrease between 2016 and 2019. 
   These authors detected a significant pressure drop at the $2.4\sigma$ level.
   However, it is worth noting that {the geometry of this occultation is grazing}.
   This may have introduced larger correlations between the pressure and the geocentric closest approach distance to Pluto's shadow axis, leading to insufficient precision to confidently support the claim of a large pressure decrease followed by a return
   in 2020 to a pressure level close to that of 2015 \citep{Sicardy2021}.

   These contrasting results highlight the need for occultation observations between 2019 and 2020 in order to better understand the behavior and evolution of Pluto's atmosphere during this time period. 
   Furthermore, while \citet{Young2021} support the presence of a pressure drop based on their analysis of the 15 August 2018 occultation, \citet{Sicardy2021} suggest that careful comparisons between measurements by independent teams should be made before drawing any conclusions on the pressure evolution.

   Observations of the 5 September 2019 occultation, which have not been reported by other teams, are presented in Section \ref{sect:obs}, followed by a description of the light-curve fitting methods in Section \ref{sect:method}. 
   These unique observations allow us to track the changes in Pluto's atmosphere during the time period between the events studied by \citet{Arimatsu2020} and \citet{Sicardy2021}.
   Results are detailed in Section \ref{sect:res}, and the pressure evolution is discussed in Section \ref{sect:dis}, including comparisons with the reanalyzed 15 August 2018 and 17 July 2019 events. 
   Conclusions and recommendations are provided in Section \ref{sect:con}.

   \section{Occultation observations}
    \label{sect:obs}

   Two observation campaigns were organized in China for occultations in 2019 (see Appendix \ref{app:cam}). 
   One occurred on {17} July 2019, which was studied by \citet{Arimatsu2020}, and the other on 5 September 2019, which is reported in the present paper for the first time.
   Due to bad weather conditions in many areas, no effective light curves were observed by our stations for the first occultation, and only two light curves were obtained for the second.

\begin{figure} 
      \centering
      \includegraphics[width=0.46\textwidth]{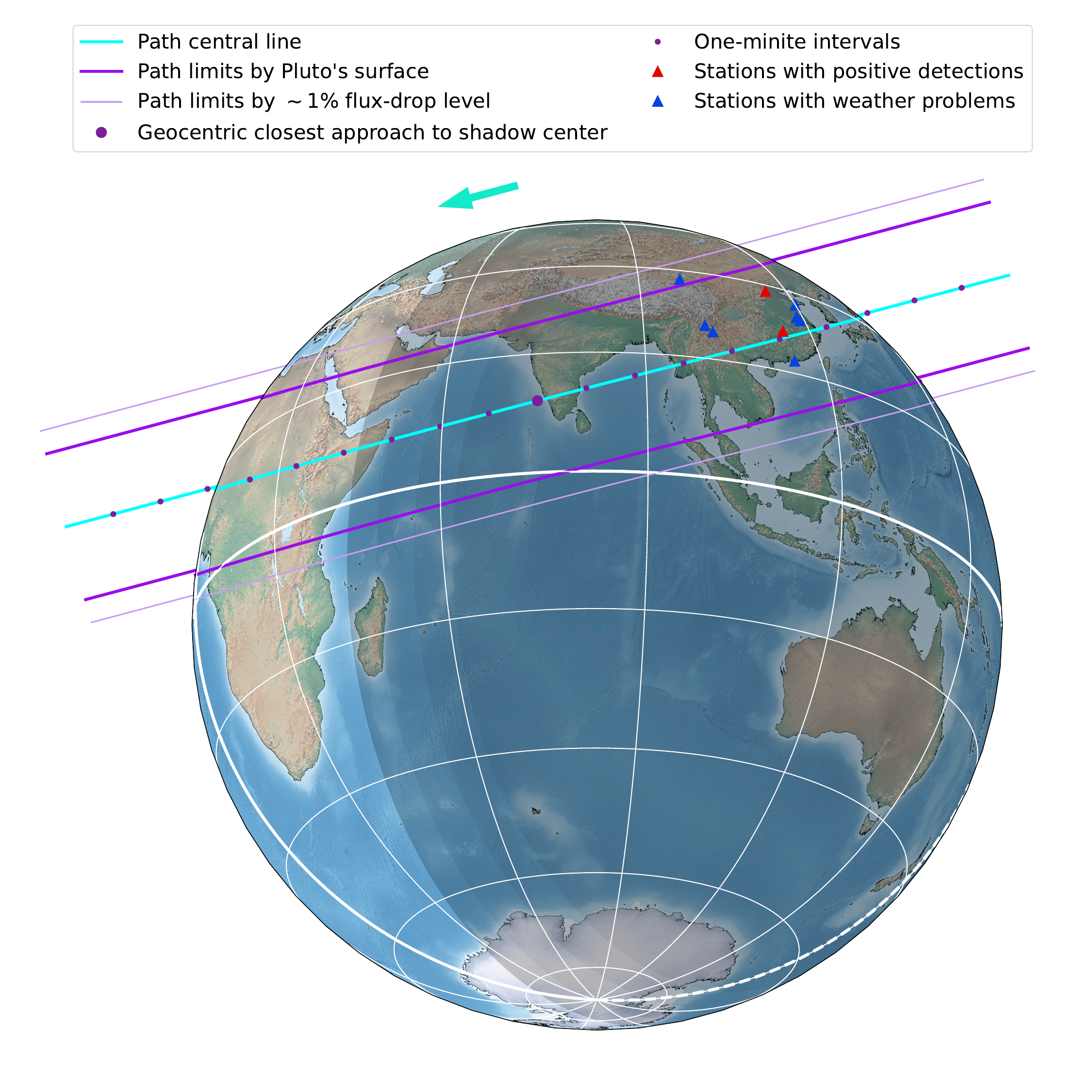}
      \caption{Reconstructed occultation map of the 5 September 2019 event.}
      \label{fig:occmap201909}
   \end{figure}

   Table \ref{tab:global201909} lists the circumstances of the 5 September 2019 event.
   Figure \ref{fig:occmap201909} presents all the observation stations and the reconstructed path of the shadow of Pluto\footnote{The occulted star is Gaia DR3 \href{https://vizier.cds.unistra.fr/viz-bin/VizieR-S?Gaia\%20DR3\%206771712487062767488}{$6771712487062767488$}, of which the astrometric and photometric parameters are obtained from {VizieR} \citep{GaiaCollaboration2022b}.} during this event.
   Table \ref{tab:pos201909} lists the circumstances of stations with positive detections.
   Their station codes are DWM and HNU.

   \begin{table}
      \centering
      \caption{Circumstances and light-curve fitting results of the 5 September 2019 event.}\label{tab:global201909}
      \setlength{\tabcolsep}{1pt}
      \begin{tabular}{ll}
         \hline\hline 
         \multicolumn{2}{c}{Occulted star} 
         \\ 
         \hline 
         Identification (Gaia DR3\tablefootmark {a}) & $6771712487062767488$ 
         \\ 
         Geocentric astrometric position & $\alpha_\txt{s} = 19^\h 29^\m 11\fs 1996$ 
         \\ 
         ~~at observational epoch (ICRF\tablefootmark {b}) & $\delta_\txt{s} = -22\degr 21'39\farcs880$ 
         \\ 
         \hline 
         \multicolumn{2}{c}{Pluto's body} 
         \\ 
         \hline 
         Mass\tablefootmark {c}, $GM_\txt{p}$ ($\km^3\cdot\s^{-2}$) & $869.6$ 
         \\ 
         Radius\tablefootmark {c}, $R_\txt{p}$ ($\km$) & $1187$
         \\ 
         \hline 
         \multicolumn{2}{c}{Pluto's atmosphere} 
         \\ 
         \hline 
         $\txt{N}_2$ molecular mass\tablefootmark {d}, $\mu$ (kg) & $4.652\times 10^{-26}$ 
         \\ 
         $\txt{N}_2$ molecular & $1.091\times 10^{-23}$ 
         \\ 
         ~~refractivity\tablefootmark {e}, $K$ ($\cm^3$)& $~~+6.282\times 10^{-26} / \lambda_\microm^2$ 
         \\ 
         Boltzmann constant\tablefootmark {f}, $k_\txt{B}$ ($\J\cdot\K^{-1}$) & $1.380649\times 10^{-23}$ 
         \\ 
         Given reference radius\tablefootmark {g}, $r_0$ (km) & $1215$ 
         \\ 
         \hline 
         \multicolumn{2}{c}{Results of atmospheric fit (with $1\sigma$ error bars)} 
         \\ 
         \hline 
         Pressure at $r_0$, $p_0$ ($\microbar$) & $6.248\pm0.30$ 
         \\ 
         Surface pressure\tablefootmark{g} at $R_\txt{p}$, $p_\txt{surf}$ ($\microbar$) & $11.478\pm0.55$ 
         \\ 
         Geocentric closest approach distance &  
         \\ 
         ~~to shadow {center}\tablefootmark {h}, $\rho_\txt{cag}$ (km) & $+3644\pm25$ 
         \\ 
         Geocentric closest approach time &  
         \\ 
         ~~to shadow {center}\tablefootmark {i}, $t_\txt{cag}$ (UTC\tablefootmark {j}) & $15\txt{:}01\txt{:}19.1\pm0.38$~s 
         \\ 
         \hline
      \end{tabular}
      \tablefoot{
         \\
         \tablefoottext{a}{\citet{GaiaCollaboration2022a}, used to derive $(\alpha_\txt{s},\delta_\txt{s})$.\\}
         \tablefoottext{b}{International Celestial Reference Frame.\\}
         \tablefoottext{c}{\citet{Stern2015}, where $G$ is the gravitational constant.\\}
         \tablefoottext{d}{Assumed to be the only constituent in the light-curve model, following, e.g., \citet{Sicardy2021}.\\}
         \tablefoottext{e}{\citet{Washburn1930}, where $\lambda_\microm$ is the wavelength expressed in microns. For stations use no filters (clear), a nominal central wavelength ($\lambda_\microm = 0.55$) was used.\\}
         \tablefoottext{f}{\citet{Newell2019}.\\}
         \tablefoottext{g}{Using a ratio $p_\txt{surf}/p_0 = 1.837$ given by the template model \citep{Meza2019,Sicardy2021}.\\}
         \tablefoottext{h}{Positive (resp. negative) values mean that the shadow {center} passes north (resp. south) of the geocenter.\\}
         \tablefoottext{i}{Timings by QHY174GPS camera are used as time references, considering their reliability and accuracy.\\}
         \tablefoottext{j}{Coordinated Universal Time.}
      }
   \end{table}

   To ensure accurate and precise timing in stellar occultations, some stations (e.g., DWM as shown in Table \ref{tab:pos201909}) were equipped with QHY174GPS cameras. 
   These cameras, manufactured by QHYCCD\footnote{\url{https://www.qhyccd.com}}, offer precise recording of observation time and location for each frame using a GPS-based function, and have been used in many stellar occultation studies \citep[e.g.,][]{Buie2020,Buie2020a,Morgado2021,Morgado2022,Pereira2023}. 
   In the light-curve fitting procedures described in Section 3.2, the time-recording offsets of the QHY174GPS cameras are fixed to zero, considering their reliability and accuracy as time references.

\begin{figure*}[h]
      \centering
      \begin{minipage}[c]{0.46\textwidth}
         \begin{subfigure}{\linewidth}
            \includegraphics[width=\linewidth]{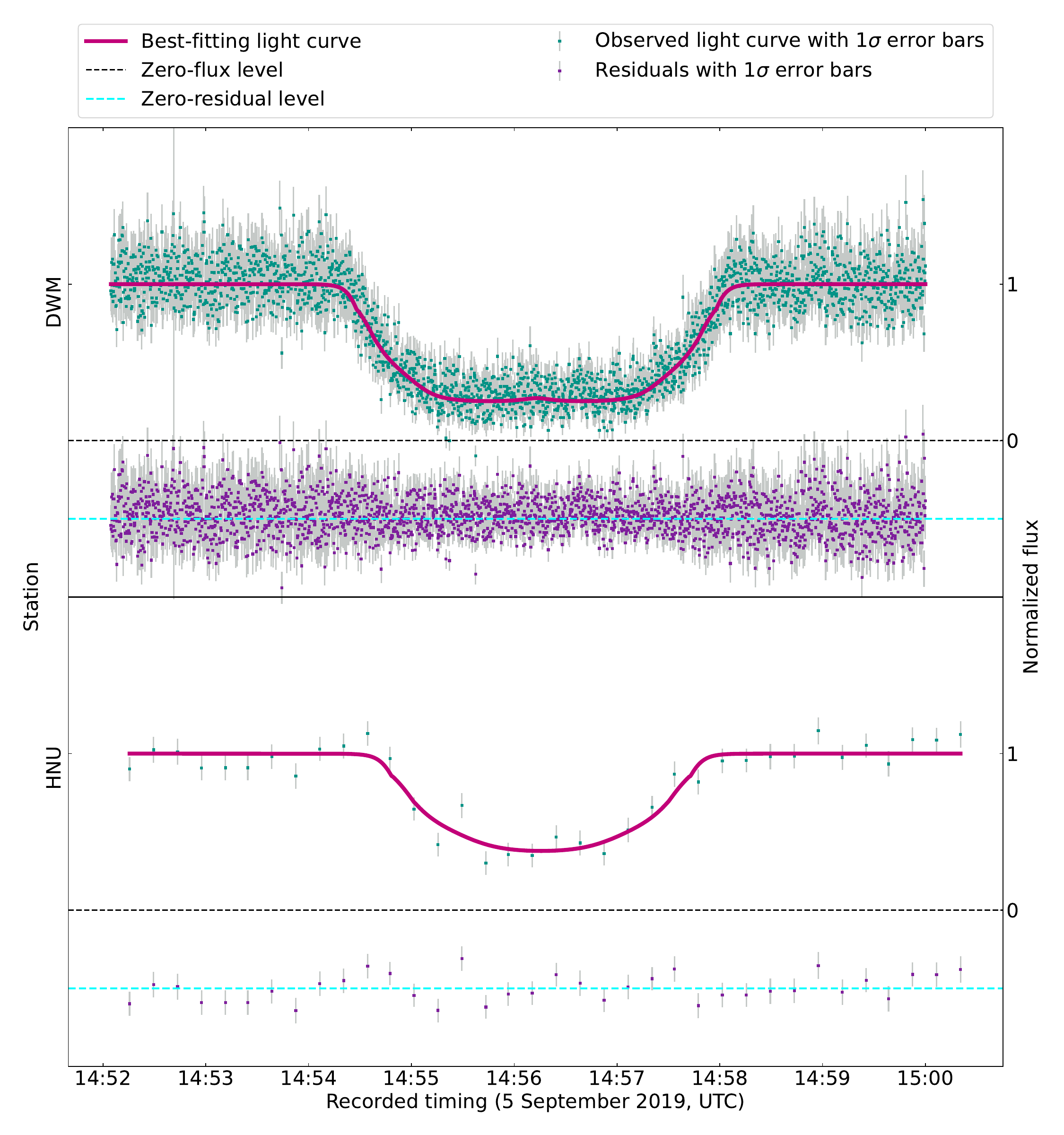}
            \caption{} 
            \label{fig:obs201909:a}
         \end{subfigure}
      \end{minipage}
      \begin{minipage}[c]{0.46\textwidth}
         \begin{subfigure}{\linewidth}
            \includegraphics[width=\linewidth]{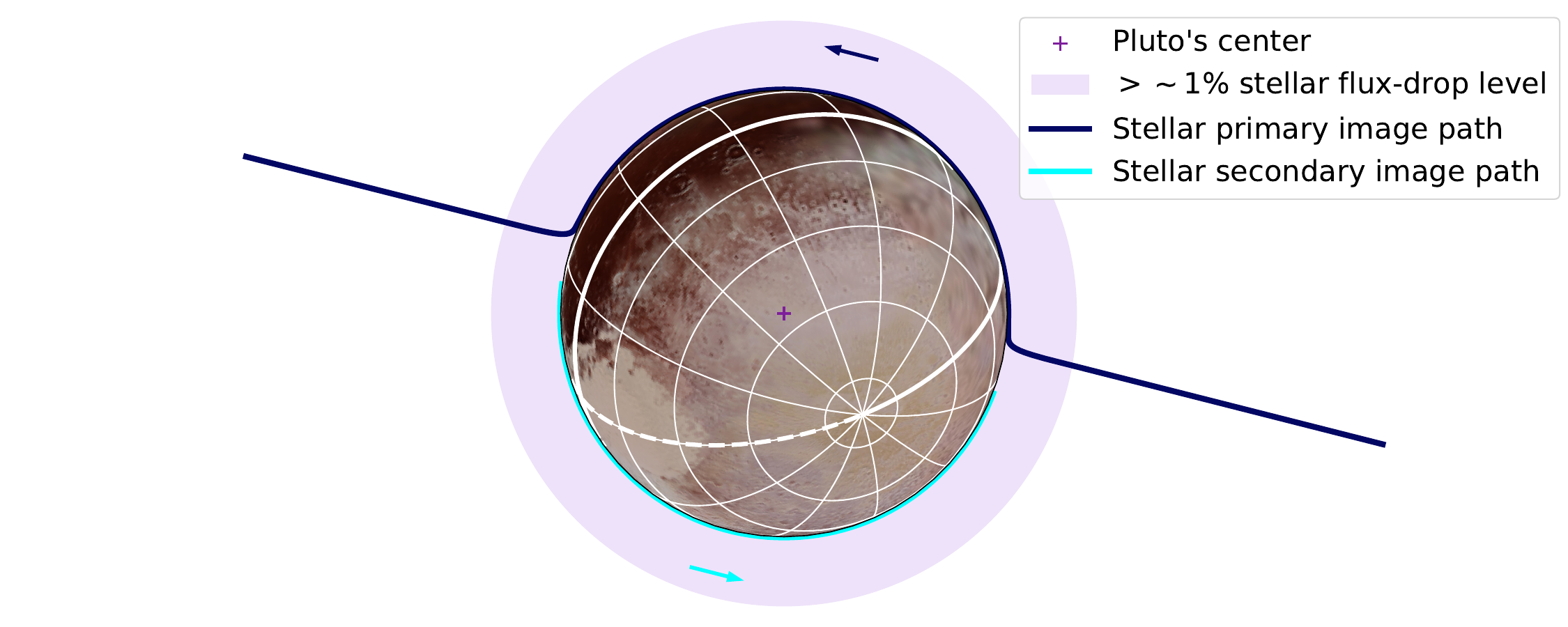}
            \caption{} 
            \label{fig:obs201909:b}
         \end{subfigure}
         \smallskip
         \begin{subfigure}{\linewidth}
            \includegraphics[width=\linewidth]{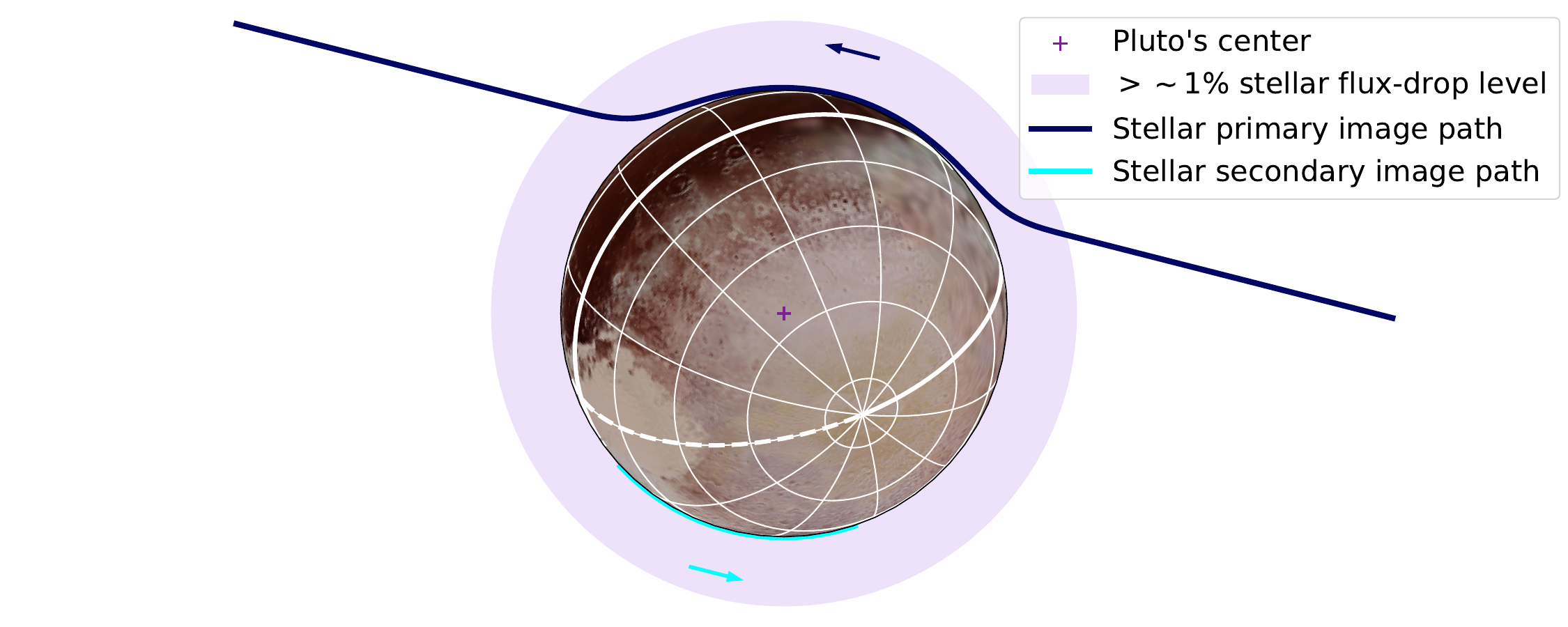}
            \caption{} 
            \label{fig:obs201909:c}
         \end{subfigure}
      \end{minipage}
      \caption{Occultation observations and the best-fitting light-curve model of the 5 September 2019 event.
      Panel (a): Observed and simultaneously fitted light curves.
      Panels (b) and (c): Reconstructed stellar paths seen by DWM and HNU, respectively.
      }
      \label{fig:obs201909}
   \end{figure*}

   All observational data were captured in the \texttt{FITS} format. 
   These data were processed using the \texttt{Tangra} occultation photometric tool\footnote{\url{http://www.hristopavlov.net/Tangra3/}} \citep{Pavlov2020} and our data-reduction code (see Appendix \ref{app:data}). 
   It was ensured that the targets and reference stars in all the images we used were not overexposed. 
   The resulting light curves from the observations, after being normalized, are presented in Figure \ref{fig:obs201909}. 
   Each data point on the light curves is represented by $f_i(t) \pm \sigma_i(t)$, where $i$ indicates the quantities associated with a specific station, $t$ represents the recorded timing of each frame, $f$ the normalized total observed flux of the occulted star and the Pluto's system, and $\sigma$ the measurement error associated with each data point.

   \section{Light-curve fitting methods}
   \label{sect:method}

   \subsection{Light-curve model}
   \label{ssect:lcmod}

   In order to simulate observed light curves, we implemented a light-curve model,
    $\phi(t;A,s,\Delta t,\Delta\tau,\Delta\rho,p_0)$, 
   which is described in Appendix \ref{app:lcmod} and is consistent with \texttt{DO15} \citep{DiasOliveira2015,Sicardy2016,Meza2019,Sicardy2021}.
   As a function of model parameters, its time-dependent Jacobian matrix was also implemented to represent the sensitivity of the model to the corresponding parameters to be estimated through fitting procedures.

   The light-curve model of a given station can be formally written as
   \begin{equation}
      \phi_i(t) = A_i \cdot \left(s_i \cdot \psi_i(t; \Delta t_i, \Delta\tau, \Delta\rho, p_0) + (1-s_i)\right),
      \label{eq:phi}
   \end{equation}
   where 
   $i$ indicates the quantities associated with the station; for further details, the reader is referred to Appendix \ref{app:lcmod}.
   Here, the reference ephemerides we use are the \texttt{NIMAv9}\footnote{\url{https://lesia.obspm.fr/lucky-star/obj.php?p=818}} asteroidal ephemeris \citep{Desmars2015,Desmars2019} for the orbit of the Pluto system barycenter with respect to the Sun, 
   the \texttt{PLU058}\footnote{\url{https://ssd.jpl.nasa.gov/ftp/eph/satellites/bsp/plu058.bsp}} satellite ephemerides \citep{Brozovic2015,Jacobson2019} for the orbit of Pluto with respect to the Pluto system barycenter, and 
   the \texttt{DE440}\footnote{\url{https://ssd.jpl.nasa.gov/ftp/eph/planets/bsp/de440.bsp}} planetary ephemerides \citep{Park2021} for the orbits of the Earth and the Sun with respect to the Solar System barycenter.
   The reference star catalog where the data of the occulted star are obtained is Gaia DR3.

   \subsection{Fitting procedure}

   The light-curve model was fitted to the normalized observed light curves simultaneously by nonlinear least squares, returning a $\chi^2$-type value of goodness-of-fit.
   The goal is to minimize the objective function given by
   \begin{equation}
      \chi^2_\txt{obs} = \sum_{i,j} \frac{(\phi_i (t_{ij}) - f_i (t_{ij}))^2}{\sigma_i^2 (t_{ij})},
   \end{equation}
   where 
   $t_{ij}$ represents the mid-exposure time of the $j$-th observation of the station $i$.

   In addition, with the used reference ephemerides and star catalog, some {a priori} information on $\Delta\rho$ can be obtained:
   \begin{equation}
      \Delta\rho = 0~\txt{km} \pm \sigma_\rho,
   \end{equation}
   where 
   the uncertainty $\sigma_\rho$ is set to $72$ km using the positional uncertainties listed in the ``orbit quality'' table of  \texttt{NIMAv9} and in Gaia DR3.
   This $\sigma_\rho$ value corresponds to about $3$ mas on the sky at the geocentric distance of Pluto.
   The {a priori} information can be treated as independent observational data and used in the model fitting, with the objective function modified as:
   \begin{equation}
      \chi^2_\txt{apr} = \chi^2_\txt{obs} + \frac{\Delta\rho^2}{\sigma_{\rho}^2}.
   \end{equation}

\begin{figure}[h]
      \centering
      \begin{subfigure}{0.92\linewidth}
         \includegraphics[width=\linewidth]{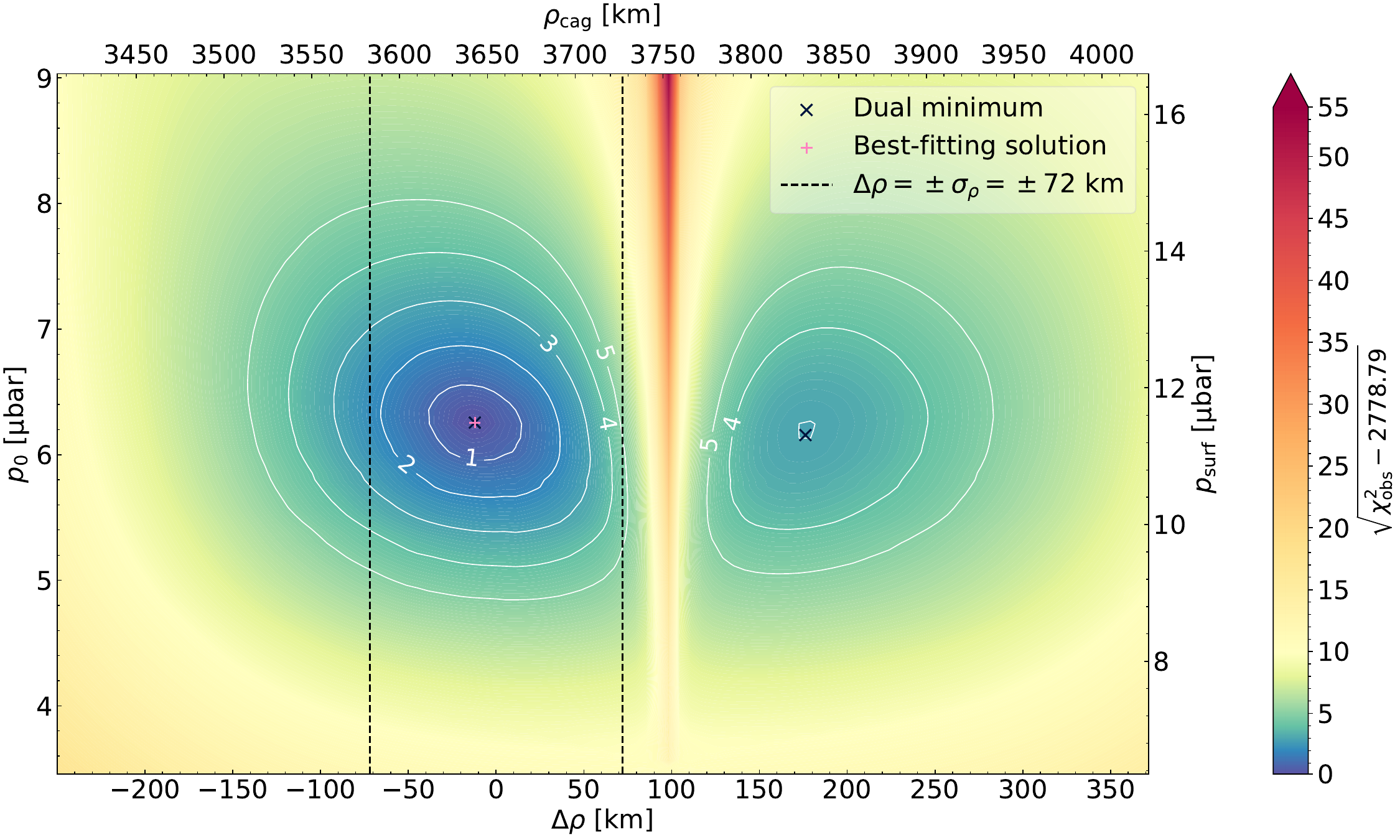}
         \caption{} 
         \label{fig:chi201909:a}
      \end{subfigure}
      \smallskip
      \begin{subfigure}{0.92\linewidth}
         \includegraphics[width=\linewidth]{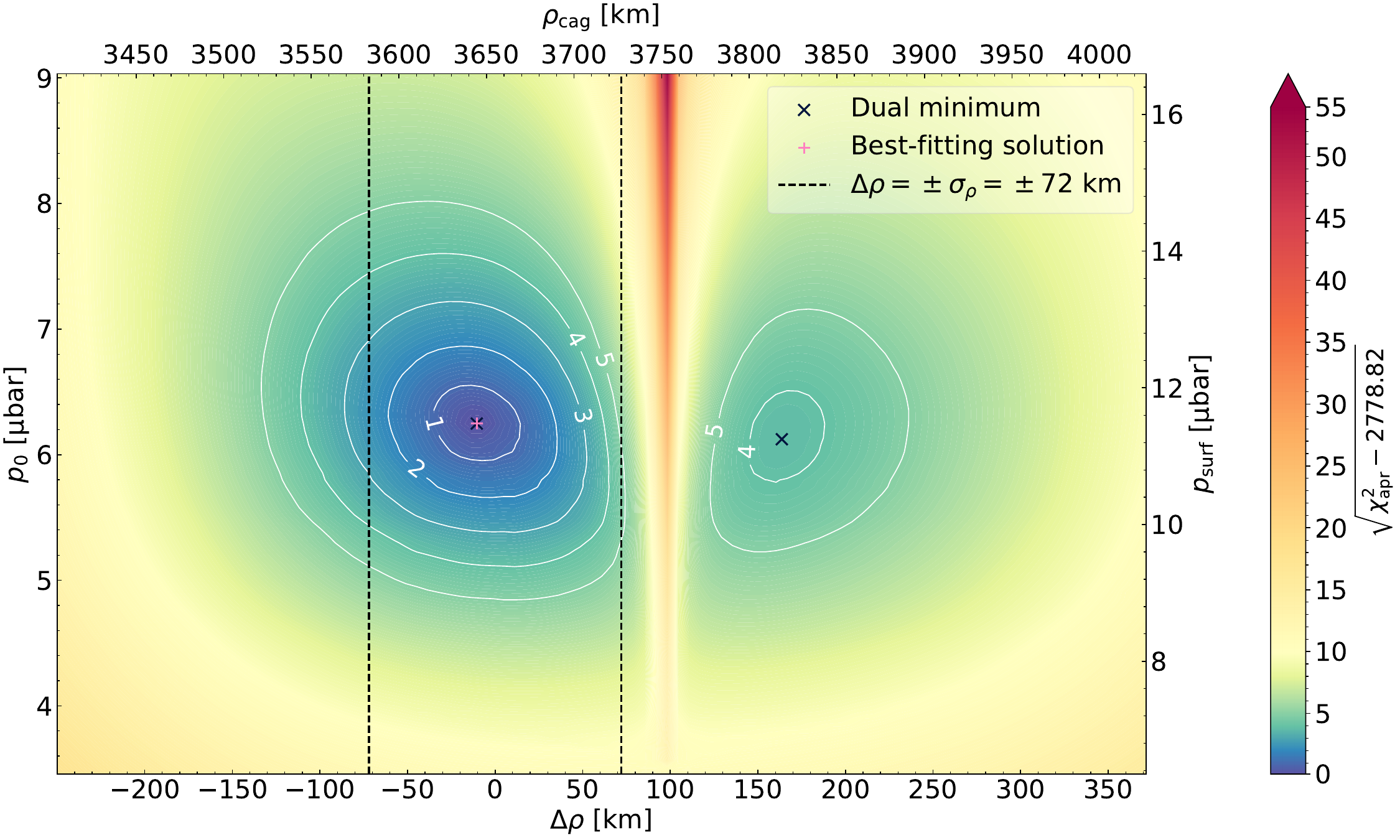}
         \caption{} 
         \label{fig:chi201909:b}
      \end{subfigure}
      \caption{The $\chi^2$ maps of the 5 September 2019 event.
      Panels (a) and (b): The $\chi^2_\txt{obs}$ and $\chi^2_\txt{apr}$ maps, respectively.
      The $\chi^2_\txt{obs}$ denotes the goodness-of-fit only using observational data, while $\chi^2_\txt{apr}$ denotes the goodness-of-fit  with the additional {a priori} information represented as the $\chi^2$ type value, $(\Delta\rho/\sigma_\rho)^2$.
      These maps are used to derive the best-fitting atmospheric pressure $p_0$ at the reference radius $r_0$ of 1215 km and the cross-track correction $\Delta\rho$ to the ephemerides, of which the {a priori} uncertainty $\sigma_\rho$ is $72$ km.
      The surface pressure $p_\txt{surf}$ and the geocentric closest approach distance $\rho_\txt{cag}$ to the shadow {center} are obtained by linear transformations of $\Delta\rho$ and $p_0$, respectively.
      The best-fitting $\chi^2_\txt{obs}$ and $\chi^2_\txt{apr}$ values per degree of freedom are $1.160$ and $1.155$, respectively.}
      \label{fig:chi201909}
   \end{figure}

   The fitting steps are as follows:
   \begin{itemize}
      \item In order to find all local minima at which a nonlinear least-squares fitting could potentially get stuck, we explored the two-parameter space $(\Delta\rho, p_0)$ by generating the variation of $\chi^2$ as a function of them.
      Figure \ref{fig:chi201909} presents such two $\chi^2$ maps, labeled (a) and (b), which are analyzed in Section \ref{sect:res}.
      The maps are generated by minimizing $\chi^2_\txt{obs}$ or $\chi^2_\txt{apr}$ at each fixed $(\Delta\rho, p_0)$ point on a regularly spaced grid.
      The Levenberg-Marquardt (LM) method, which is implemented in the \texttt{LMFIT} package\footnote{\url{https://lmfit.github.io/lmfit-py/}}, was used in each fitting procedure.
      The free parameters to be adjusted are $\Delta\tau$\footnote{As a technical note, if no equipment for reliable time references is used, $\Delta\tau$ should be fixed as zero, avoiding fitting problems caused by its linear relationship with $t_i$.}, $\Delta t_i$ of any station with no reliable time reference system like QHY174GPS, and $s_i$ and $A_i$ of each station.
      \item For a more accurate best-fitting solution for $(\Delta\rho, p_0)$, the LM method is used again, with $\Delta\rho$ and $p_0$ adjusted with initial guesses located at all known local minima of each $\chi^2$ map.
      \item Each $\chi^2$ map, which provides information about the quality of the fit, is used to define confidence limits based on constant $\chi^2$ boundaries \citep{Press2007}.
   \end{itemize}

   \section{Results}
   \label{sect:res}

   Figure \ref{fig:chi201909:a} shows the $\chi^2_\txt{obs}$ map for the 5 September 2019 occultation. 
   Two local minima are observed.
   However, considering the significant $\chi^2$ difference of $9$ between the two local minima, the global minimum is more likely to be the correct solution. 
   In addition, the $\Delta\rho$ value at the global minimum is more consistent with the \texttt{NIMAv9} solution, $\Delta\rho=0$ km, at the $0.16~\sigma_\rho$ level, compared with the other local one at the $2.44~\sigma_\rho$ level. 

   In an effort to mitigate or at least further weaken the presence of multiple local minima, we calculated the $\chi^2_\txt{apr}$ map by adding the $\chi^2$-type value of the {a priori} information, ${(\Delta\rho/\sigma_{\rho})^2}$, into the $\chi^2_{\txt{obs}}$ map. 
   Figure \ref{fig:chi201909:b} presents the results, which show that two local minima are still present, but with {a $\chi^2$ difference of about 14.5, which is larger than that of the $\chi^2_\text{obs}$ map}. 
   Therefore,  the global minimum is confidently accepted as the solution for $(\rho_{\txt{cag}}, p_{\txt{surf}})$, as provided in Table \ref{tab:global201909}.

   Moreover, Figure \ref{fig:chi201909} presents the consistency of our derived $p_{\txt{surf}}$ across the two different local minima.
   Our findings demonstrate that the specific choice of local minima does not significantly affect the value of $p_{\txt{surf}}$, further supporting the reliability of our solution for $p_{\txt{surf}}$.

   \section{Pressure evolution}
   \label{sect:dis}

   \subsection{Comparisons and necessary reanalyses of historical events}

\begin{figure} 
      \centering
      \includegraphics[width=\linewidth]{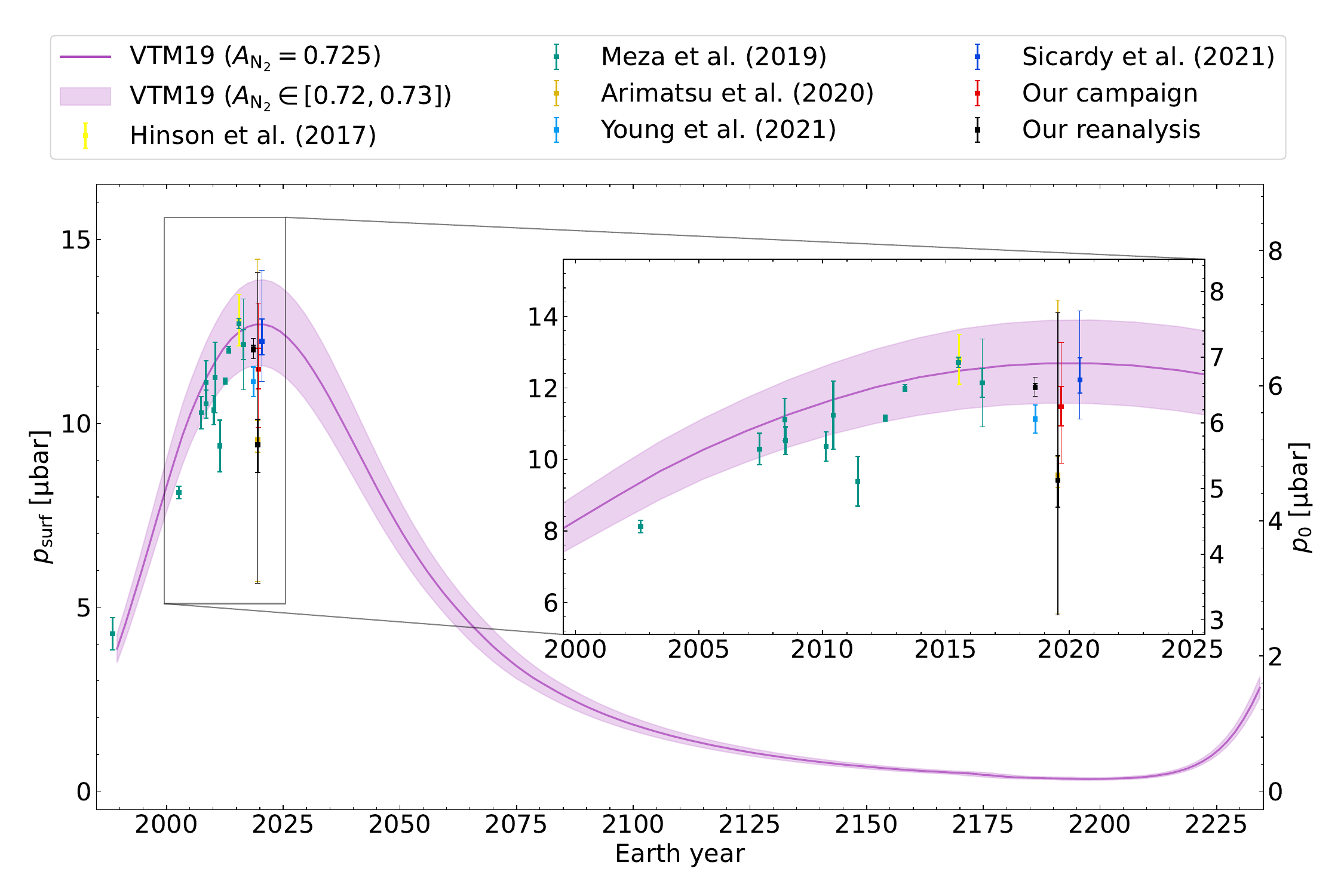}
      \caption{Pressure evolution of Pluto over a 248-year heliocentric orbital period predicted by the {VTM19}, along with the measured pressures with $1\sigma$ error bars. {The $3\sigma$ error bars of our new measurements and of the previous measurements of the 19 July 2016, 17 July 2019, and 6 June 2020 events are also presented using the published $\chi^2$ maps.} $A_\txt{N_2}$ denotes the albedo of nitrogen ice.}
      \label{fig:lmd}
   \end{figure}

   In Figure \ref{fig:lmd}, the red plot represents our $p_\txt{surf}$ measurement from the 5 September 2019 occultation. 
   We also include other published measurements \citep{Hinson2017,Meza2019,Arimatsu2020,Young2021,Sicardy2021} and the pressure evolution predicted by the VTM19 in order to provide a comprehensive view of the pressure variations on Pluto.

   To avoid potential inconsistencies arising from different analysis methods, as discussed by \citet{Sicardy2021}, we reanalyzed the 15 August 2018 event studied by \citet{Young2021} using the {IXON} observational data of \citet{SilvaCabrera2022}. 
   The derived pressure measurement presented in Appendix \ref{sapp:2018} is ${12.027}_{-0.08}^{+0.09}~\txt{\upmu bar}$.
   In addition, we also reanalyze the 17 July 2019 event in Appendix \ref{sapp:2019}, deriving a pressure of $p_\txt{surf} = 9.421_{-0.75}^{+0.68}~\microbar$, which is similar to that of \citet{Arimatsu2020}, of  ${9.56}_{-0.34}^{+0.52}~\microbar$. 
   This similarity is expected because the same \texttt{DO15} method is used. 
   As this same method is used by \citet{Meza2019} and \citet{Sicardy2021}, their pressure measurements, along with that of \citet{Arimatsu2020}, can be fully compared with our new ones.
   Both the remeasurements are plotted in {black} in Figure \ref{fig:lmd}.

   {The pressure measurement from the 5 September 2019 event shows alignments with those from the 19 July 2016, 15 August 2018, and 6 June 2020 events within their combined $1\sigma$ levels.
   Our new measurement from the 15 August 2018 event does not show the significant pressure drop previously reported by \citet{Young2021}.
   The previously reported pressure drop between the 19 July 2016 and 17 July 2019 events is still detected at the same level as in \citet{Arimatsu2020}.}

   \subsection{Discussion on pressure variations}

   {While the VTM19 remains, on average, applicable and capable of predicting the main atmospheric behavior during the observed years, there are also two probable V-shaped pressure variations observed from 2010 to 2015 and from 2015 to 2020, especially when considering the measurements from the 4 June 2011 and 17 July 2019 events.}
   {These V-shaped variations suggest} the presence of additional factors that have not been accounted for. 
   Specifically, short-term changes in Pluto's surface ices and their interaction with the atmosphere are likely contributing to the variation. 
   Moreover, spectral monitoring of the surface composition has revealed some short-term changes in the ices over several Earth years \citep[e.g.,][]{Grundy2014,Lellouch2022,Holler2022}.

   {However, the validity of the V-shaped variations is debatable given the stellar faintness of the 4 June 2011 event and the grazing single-chord geometry of the 17 July 2019 event.
   If the debatable measurement from the 17 July 2019 event were discarded, no significant changes would be observed between 2016 and 2020.   
   This more likely supports the plateau phase since 2015 predicted by the VTM19.}
   {In order} to better understand the relationship between these factors, further observations using multiple observational techniques (occultation, spectroscopy, and photometry) are required, as well as simulations with a refined VTM.

   \section{Conclusions}
   \label{sect:con}

   The unique observations of the 5 September 2019 occultation provide a surface pressure of $p_\txt{surf} = 11.478\pm 0.55~\microbar$. 
   In order to avoid potential method inconsistencies in comparing with historical pressure measurements \citep{Sicardy2021}, we also reanalyzed the 15 August 2018 and {17 July} 2019 events based on publicly available data \citep{SilvaCabrera2022,Arimatsu2020}.
   All measurements are presented in Figure \ref{fig:lmd}.

   The VTM19 remains applicable on average. 
   {In addition, we also observed unaccounted-for V-shaped pressure variations with the previously reported pressure drop being a part of these variations; however, these variations are debatable. }
   To better understand all significant pressure variations of Pluto, continuous pressure monitoring through occultation observations is essential where possible.
   Also, simultaneous and frequent spectroscopic and photometric monitoring of  changes to its surface ice are important, as such comprehensive monitoring will provide more short-term and long-term evolution constraints of Pluto's interacting atmosphere and surface.

   \begin{acknowledgements}

      We acknowledge Bruno Sicardy for his useful comments that helped improving this manuscript.
      This work has been supported by the National Natural Science Foundation of China (Grant Nos. 12203105 and 12103091).
      We acknowledge the science research grants from the China Manned Space Project with NO.CMS-CSST-2021-A12 and NO.CMS-CSST-2021-B10.
      This research has made use of data from the 40 cm DOB telescope at the Dawei Mountain Observatory of Hunan Astronomical Association and from the 50 cm RC telescope at the Observatory of Hebei Normal University.
      We ackonwledge the support of Chinese amateur astronomers from Hunan Astronomical Association, Nanjing Amateur Astronomers Association, and Shenzhen Astronomical Observatory Team.

   \end{acknowledgements}

   \bibliographystyle{aa}
   \bibliography{pluto_occ_2019.bib}

\begin{thebibliography}{49}
\expandafter\ifx\csname natexlab\endcsname\relax\def\natexlab#1{#1}\fi

\bibitem[{{Arimatsu} {et~al.}(2020){Arimatsu}, {Hashimoto}, {Kagitani},
  {Sakanoi}, {Kasaba}, {Ohsawa}, \& {Urakawa}}]{Arimatsu2020}
{Arimatsu}, K., {Hashimoto}, G.~L., {Kagitani}, M., {et~al.} 2020, \aap, 638,
  L5

\bibitem[{{Bertrand} \& {Forget}(2016)}]{Bertrand2016}
{Bertrand}, T. \& {Forget}, F. 2016, \nat, 540, 86

\bibitem[{{Bertrand} {et~al.}(2018){Bertrand}, {Forget}, {Umurhan}, {Grundy},
  {Schmitt}, {Protopapa}, {Zangari}, {White}, {Schenk}, {Singer}, {Stern},
  {Weaver}, {Young}, {Ennico}, \& {Olkin}}]{Bertrand2018}
{Bertrand}, T., {Forget}, F., {Umurhan}, O.~M., {et~al.} 2018, \icarus, 309,
  277

\bibitem[{{Bertrand} {et~al.}(2019){Bertrand}, {Forget}, {Umurhan}, {Moore},
  {Young}, {Protopapa}, {Grundy}, {Schmitt}, {Dhingra}, {Binzel}, {Earle},
  {Cruikshank}, {Stern}, {Weaver}, {Ennico}, {Olkin}, \& {New Horizons Science
  Team}}]{Bertrand2019}
{Bertrand}, T., {Forget}, F., {Umurhan}, O.~M., {et~al.} 2019, \icarus, 329,
  148

\bibitem[{{Bosh} {et~al.}(2015){Bosh}, {Person}, {Levine}, {Zuluaga},
  {Zangari}, {Gulbis}, {Schaefer}, {Dunham}, {Babcock}, {Davis}, {Pasachoff},
  {Rojo}, {Servajean}, {F{\"o}rster}, {Oswalt}, {Batcheldor}, {Bell}, {Bird},
  {Fey}, {Fulwider}, {Geisert}, {Hastings}, {Keuhler}, {Mizusawa}, {Solenski},
  \& {Watson}}]{Bosh2015}
{Bosh}, A.~S., {Person}, M.~J., {Levine}, S.~E., {et~al.} 2015, \icarus, 246,
  237

\bibitem[{{Brosch}(1995)}]{Brosch1995}
{Brosch}, N. 1995, \mnras, 276, 571

\bibitem[{{Brozovi{\'c}} {et~al.}(2015){Brozovi{\'c}}, {Showalter}, {Jacobson},
  \& {Buie}}]{Brozovic2015}
{Brozovi{\'c}}, M., {Showalter}, M.~R., {Jacobson}, R.~A., \& {Buie}, M.~W.
  2015, \icarus, 246, 317

\bibitem[{{Buie} {et~al.}(2020{\natexlab{a}}){Buie}, {Leiva}, {Keller},
  {Desmars}, {Sicardy}, {Kavelaars}, {Bridges}, {Weryk}, {Herald}, {Haley},
  {Strauss}, {Wilde}, {Baker}, {Conway}, {Dean}, {Dunham}, {Estes}, {Fiechter},
  {Givot}, {Glibbery}, {Gowe}, {Hayman}, {Ireland}, {Kehrli}, {Moore},
  {MacDonald}, {McCrystal}, {Mendoza}, {Palmquist}, {Rennau}, {Schar},
  {Swanson}, {Terris}, {Werts}, \& {Wise}}]{Buie2020}
{Buie}, M.~W., {Leiva}, R., {Keller}, J.~M., {et~al.} 2020{\natexlab{a}}, \aj,
  159, 230

\bibitem[{{Buie} {et~al.}(2020{\natexlab{b}}){Buie}, {Porter}, {Tamblyn},
  {Terrell}, {Parker}, {Baratoux}, {Kaire}, {Leiva}, {Verbiscer}, {Zangari},
  {Colas}, {Diop}, {Samaniego}, {Wasserman}, {Benecchi}, {Caspi}, {Gwyn},
  {Kavelaars}, {Ocampo Ur{\'\i}a}, {Rabassa}, {Skrutskie}, {Soto}, {Tanga},
  {Young}, {Stern}, {Andersen}, {Arango P{\'e}rez}, {Arredondo}, {Artola},
  {B{\^a}}, {Ballet}, {Blank}, {Bop}, {Bosh}, {Camino L{\'o}pez}, {Carter},
  {Castro-Chac{\'o}n}, {Caycedo Desprez}, {Caycedo Guerra}, {Conard},
  {Dauvergne}, {Dean}, {Dean}, {Desmars}, {Dieng}, {Bousso Dieng}, {Diouf},
  {Dorego}, {Dunham}, {Dunham}, {Durantini Luca}, {Edwards}, {Erasmus}, {Faye},
  {Faye}, {Ferrario}, {Ferrell}, {Finley}, {Fraser}, {Friedli}, {Galvez Serna},
  {Garcia-Migani}, {Genade}, {Getrost}, {Gil-Hutton}, {Gimeno}, {Golub},
  {Gonz{\'a}lez Murillo}, {Grusin}, {Gurovich}, {Hanna}, {Henn}, {Hinton},
  {Hughes}, {Josephs}, {Joya}, {Kammer}, {Keeney}, {Keller}, {Kramer},
  {Levine}, {Lisse}, {Lovell}, {Mackie}, {Makarchuk}, {Manzano}, {Mbaye},
  {Mbaye}, {Melia}, {Moreno}, {Moss}, {Ndaiye}, {Ndiaye}, {Nelson}, {Olkin},
  {Olsen}, {Ospina Moreno}, {Pasachoff}, {Pereyra}, {Person}, {Pinz{\'o}n},
  {Pulver}, {Quintero}, {Regester}, {Resnick}, {Reyes-Ruiz}, {Rolfsmeier},
  {Ruhland}, {Salmon}, {Santos-Sanz}, {Santucho}, {Sep{\'u}lveda Ni{\~n}o},
  {Sickafoose}, {Silva}, {Singer}, {Skipper}, {Slivan}, {Smith}, {Spagnotto},
  {Stephens}, {Strabala}, {Tamayo}, {Throop}, {Torres Ca{\~n}as}, {Toure},
  {Traore}, {Tsang}, {Turner}, {Vanegas}, {Venable}, {Wilson}, {Zuluaga}, \&
  {Zuluaga}}]{Buie2020a}
{Buie}, M.~W., {Porter}, S.~B., {Tamblyn}, P., {et~al.} 2020{\natexlab{b}},
  \aj, 159, 130

\bibitem[{{Desmars} {et~al.}(2015){Desmars}, {Camargo}, {Braga-Ribas},
  {Vieira-Martins}, {Assafin}, {Vachier}, {Colas}, {Ortiz}, {Duffard},
  {Morales}, {Sicardy}, {Gomes-J{\'u}nior}, \& {Benedetti-Rossi}}]{Desmars2015}
{Desmars}, J., {Camargo}, J.~I.~B., {Braga-Ribas}, F., {et~al.} 2015, A\&A,
  584, A96

\bibitem[{{Desmars} {et~al.}(2019){Desmars}, {Meza}, {Sicardy}, {Assafin},
  {Camargo}, {Braga-Ribas}, {Benedetti-Rossi}, {Dias-Oliveira}, {Morgado},
  {Gomes-J{\'u}nior}, {Vieira-Martins}, {Behrend}, {Ortiz}, {Duffard},
  {Morales}, \& {Santos Sanz}}]{Desmars2019}
{Desmars}, J., {Meza}, E., {Sicardy}, B., {et~al.} 2019, \aap, 625, A43

\bibitem[{{Dias-Oliveira} {et~al.}(2015){Dias-Oliveira}, {Sicardy}, {Lellouch},
  {Vieira-Martins}, {Assafin}, {Camargo}, {Braga-Ribas}, {Gomes-J{\'u}nior},
  {Benedetti-Rossi}, {Colas}, {Decock}, {Doressoundiram}, {Dumas}, {Emilio},
  {Fabrega Polleri}, {Gil-Hutton}, {Gillon}, {Girard}, {Hau}, {Ivanov},
  {Jehin}, {Lecacheux}, {Leiva}, {Lopez-Sisterna}, {Mancini}, {Manfroid},
  {Maury}, {Meza}, {Morales}, {Nagy}, {Opitom}, {Ortiz}, {Pollock}, {Roques},
  {Snodgrass}, {Soulier}, {Thirouin}, {Vanzi}, {Widemann}, {Reichart},
  {LaCluyze}, {Haislip}, {Ivarsen}, {Dominik}, {J{\o}rgensen}, \&
  {Skottfelt}}]{DiasOliveira2015}
{Dias-Oliveira}, A., {Sicardy}, B., {Lellouch}, E., {et~al.} 2015, \apj, 811,
  53

\bibitem[{{Elliot} {et~al.}(2003){Elliot}, {Ates}, {Babcock}, {Bosh}, {Buie},
  {Clancy}, {Dunham}, {Eikenberry}, {Hall}, {Kern}, {Leggett}, {Levine},
  {Moon}, {Olkin}, {Osip}, {Pasachoff}, {Penprase}, {Person}, {Qu}, {Rayner},
  {Roberts}, {Salyk}, {Souza}, {Stone}, {Taylor}, {Tholen}, {Thomas-Osip},
  {Ticehurst}, \& {Wasserman}}]{Elliot2003a}
{Elliot}, J.~L., {Ates}, A., {Babcock}, B.~A., {et~al.} 2003, \nat, 424, 165

\bibitem[{{Elliot} {et~al.}(1989){Elliot}, {Dunham}, {Bosh}, {Slivan}, {Young},
  {Wasserman}, \& {Millis}}]{Elliot1989}
{Elliot}, J.~L., {Dunham}, E.~W., {Bosh}, A.~S., {et~al.} 1989, \icarus, 77,
  148

\bibitem[{{Forget} {et~al.}(2017){Forget}, {Bertrand}, {Vangvichith},
  {Leconte}, {Millour}, \& {Lellouch}}]{Forget2017}
{Forget}, F., {Bertrand}, T., {Vangvichith}, M., {et~al.} 2017, \icarus, 287,
  54

\bibitem[{{Gaia Collaboration}(2022)}]{GaiaCollaboration2022b}
{Gaia Collaboration}. 2022, VizieR Online Data Catalog, I/355

\bibitem[{{Gaia Collaboration} {et~al.}(2022){Gaia Collaboration}, {Vallenari},
  {Brown}, {Prusti}, {de Bruijne}, {Arenou}, {Babusiaux}, {Biermann},
  {Creevey}, {Ducourant}, {Evans}, {Eyer}, {Guerra}, {Hutton}, {Jordi},
  {Klioner}, {Lammers}, {Lindegren}, {Luri}, {Mignard}, {Panem}, {Pourbaix},
  {Randich}, {Sartoretti}, {Soubiran}, {Tanga}, {Walton}, {Bailer-Jones},
  {Bastian}, {Drimmel}, {Jansen}, {Katz}, {Lattanzi}, {van Leeuwen}, {Bakker},
  {Cacciari}, {Casta{\~n}eda}, {De Angeli}, {Fabricius}, {Fouesneau},
  {Fr{\'e}mat}, {Galluccio}, {Guerrier}, {Heiter}, {Masana}, {Messineo},
  {Mowlavi}, {Nicolas}, {Nienartowicz}, {Pailler}, {Panuzzo}, {Riclet}, {Roux},
  {Seabroke}, {Sordo{\o}rcit}, {Th{\'e}venin}, {Gracia-Abril}, {Portell},
  {Teyssier}, {Altmann}, {Andrae}, {Audard}, {Bellas-Velidis}, {Benson},
  {Berthier}, {Blomme}, {Burgess}, {Busonero}, {Busso}, {C{\'a}novas}, {Carry},
  {Cellino}, {Cheek}, {Clementini}, {Damerdji}, {Davidson}, {de Teodoro},
  {Nu{\~n}ez Campos}, {Delchambre}, {Dell'Oro}, {Esquej},
  {Fern{\'a}ndez-Hern{\'a}ndez}, {Fraile}, {Garabato}, {Garc{\'\i}a-Lario},
  {Gosset}, {Haigron}, {Halbwachs}, {Hambly}, {Harrison}, {Hern{\'a}ndez},
  {Hestroffer}, {Hodgkin}, {Holl}, {Jan{\ss}en}, {Jevardat de Fombelle},
  {Jordan}, {Krone-Martins}, {Lanzafame}, {L{\"o}ffler}, {Marchal}, {Marrese},
  {Moitinho}, {Muinonen}, {Osborne}, {Pancino}, {Pauwels}, {Recio-Blanco},
  {Reyl{\'e}}, {Riello}, {Rimoldini}, {Roegiers}, {Rybizki}, {Sarro}, {Siopis},
  {Smith}, {Sozzetti}, {Utrilla}, {van Leeuwen}, {Abbas}, {{\'A}brah{\'a}m},
  {Abreu Aramburu}, {Aerts}, {Aguado}, {Ajaj}, {Aldea-Montero}, {Altavilla},
  {{\'A}lvarez}, {Alves}, {Anders}, {Anderson}, {Anglada Varela}, {Antoja},
  {Baines}, {Baker}, {Balaguer-N{\'u}{\~n}ez}, {Balbinot}, {Balog}, {Barache},
  {Barbato}, {Barros}, {Barstow}, {Bartolom{\'e}}, {Bassilana}, {Bauchet},
  {Becciani}, {Bellazzini}, {Berihuete}, {Bernet}, {Bertone}, {Bianchi},
  {Binnenfeld}, {Blanco-Cuaresma}, {Blazere}, {Boch}, {Bombrun}, {Bossini},
  {Bouquillon}, {Bragaglia}, {Bramante}, {Breedt}, {Bressan}, {Brouillet},
  {Brugaletta}, {Bucciarelli}, {Burlacu}, {Butkevich}, {Buzzi}, {Caffau},
  {Cancelliere}, {Cantat-Gaudin}, {Carballo}, {Carlucci}, {Carnerero},
  {Carrasco}, {Casamiquela}, {Castellani}, {Castro-Ginard}, {Chaoul},
  {Charlot}, {Chemin}, {Chiaramida}, {Chiavassa}, {Chornay}, {Comoretto},
  {Contursi}, {Cooper}, {Cornez}, {Cowell}, {Crifo}, {Cropper}, {Crosta},
  {Crowley}, {Dafonte}, {Dapergolas}, {David}, {David}, {de Laverny}, {De
  Luise}, {De March}, {De Ridder}, {de Souza}, {de Torres}, {del Peloso}, {del
  Pozo}, {Delbo}, {Delgado}, {Delisle}, {Demouchy}, {Dharmawardena}, {Di
  Matteo}, {Diakite}, {Diener}, {Distefano}, {Dolding}, {Edvardsson}, {Enke},
  {Fabre}, {Fabrizio}, {Faigler}, {Fedorets}, {Fernique}, {Fienga}, {Figueras},
  {Fournier}, {Fouron}, {Fragkoudi}, {Gai}, {Garcia-Gutierrez},
  {Garcia-Reinaldos}, {Garc{\'\i}a-Torres}, {Garofalo}, {Gavel}, {Gavras},
  {Gerlach}, {Geyer}, {Giacobbe}, {Gilmore}, {Girona}, {Giuffrida}, {Gomel},
  {Gomez}, {Gonz{\'a}lez-N{\'u}{\~n}ez}, {Gonz{\'a}lez-Santamar{\'\i}a},
  {Gonz{\'a}lez-Vidal}, {Granvik}, {Guillout}, {Guiraud},
  {Guti{\'e}rrez-S{\'a}nchez}, {Guy}, {Hatzidimitriou}, {Hauser}, {Haywood},
  {Helmer}, {Helmi}, {Sarmiento}, {Hidalgo}, {Hilger}, {H{\l}adczuk}, {Hobbs},
  {Holland}, {Huckle}, {Jardine}, {Jasniewicz}, {Jean-Antoine Piccolo},
  {Jim{\'e}nez-Arranz}, {Jorissen}, {Juaristi Campillo}, {Julbe}, {Karbevska},
  {Kervella}, {Khanna}, {Kontizas}, {Kordopatis}, {Korn}, {K{\'o}sp{\'a}l},
  {Kostrzewa-Rutkowska}, {Kruszy{\'n}ska}, {Kun}, {Laizeau}, {Lambert},
  {Lanza}, {Lasne}, {Le Campion}, {Lebreton}, {Lebzelter}, {Leccia}, {Leclerc},
  {Lecoeur-Taibi}, {Liao}, {Licata}, {Lindstr{\o}m}, {Lister}, {Livanou},
  {Lobel}, {Lorca}, {Loup}, {Madrero Pardo}, {Magdaleno Romeo}, {Managau},
  {Mann}, {Manteiga}, {Marchant}, {Marconi}, {Marcos}, {Marcos Santos},
  {Mar{\'\i}n Pina}, {Marinoni}, {Marocco}, {Marshall}, {Polo},
  {Mart{\'\i}n-Fleitas}, {Marton}, {Mary}, {Masip}, {Massari},
  {Mastrobuono-Battisti}, {Mazeh}, {McMillan}, {Messina}, {Michalik}, {Millar},
  {Mints}, {Molina}, {Molinaro}, {Moln{\'a}r}, {Monari}, {Mongui{\'o}},
  {Montegriffo}, {Montero}, {Mor}, {Mora}, {Morbidelli}, {Morel}, {Morris},
  {Muraveva}, {Murphy}, {Musella}, {Nagy}, {Noval}, {Oca{\~n}a}, {Ogden},
  {Ordenovic}, {Osinde}, {Pagani}, {Pagano}, {Palaversa}, {Palicio},
  {Pallas-Quintela}, {Panahi}, {Payne-Wardenaar}, {Pe{\~n}alosa Esteller},
  {Penttil{\"a}}, {Pichon}, {Piersimoni}, {Pineau}, {Plachy}, {Plum}, {Poggio},
  {Pr{\v{s}}a}, {Pulone}, {Racero}, {Ragaini}, {Rainer}, {Raiteri}, {Rambaux},
  {Ramos}, {Ramos-Lerate}, {Re Fiorentin}, {Regibo}, {Richards}, {Rios Diaz},
  {Ripepi}, {Riva}, {Rix}, {Rixon}, {Robichon}, {Robin}, {Robin}, {Roelens},
  {Rogues}, {Rohrbasser}, {Romero-G{\'o}mez}, {Rowell}, {Royer}, {Ruz Mieres},
  {Rybicki}, {Sadowski}, {S{\'a}ez N{\'u}{\~n}ez}, {Sagrist{\`a} Sell{\'e}s},
  {Sahlmann}, {Salguero}, {Samaras}, {Sanchez Gimenez}, {Sanna},
  {Santove{\~n}a}, {Sarasso}, {Schultheis}, {Sciacca}, {Segol}, {Segovia},
  {S{\'e}gransan}, {Semeux}, {Shahaf}, {Siddiqui}, {Siebert}, {Siltala},
  {Silvelo}, {Slezak}, {Slezak}, {Smart}, {Snaith}, {Solano}, {Solitro},
  {Souami}, {Souchay}, {Spagna}, {Spina}, {Spoto}, {Steele},
  {Steidelm{\"u}ller}, {Stephenson}, {S{\"u}veges}, {Surdej}, {Szabados},
  {Szegedi-Elek}, {Taris}, {Taylo}, {Teixeira}, {Tolomei}, {Tonello}, {Torra},
  {Torra}, {Torralba Elipe}, {Trabucchi}, {Tsounis}, {Turon}, {Ulla}, {Unger},
  {Vaillant}, {van Dillen}, {van Reeven}, {Vanel}, {Vecchiato}, {Viala},
  {Vicente}, {Voutsinas}, {Weiler}, {Wevers}, {Wyrzykowski}, {Yoldas}, {Yvard},
  {Zhao}, {Zorec}, {Zucker}, \& {Zwitter}}]{GaiaCollaboration2022a}
{Gaia Collaboration}, {Vallenari}, A., {Brown}, A.~G.~A., {et~al.} 2022, arXiv
  e-prints, arXiv:2208.00211

\bibitem[{{Grundy} {et~al.}(2014){Grundy}, {Olkin}, {Young}, \&
  {Holler}}]{Grundy2014}
{Grundy}, W.~M., {Olkin}, C.~B., {Young}, L.~A., \& {Holler}, B.~J. 2014,
  \icarus, 235, 220

\bibitem[{{Gulbis} {et~al.}(2015){Gulbis}, {Emery}, {Person}, {Bosh},
  {Zuluaga}, {Pasachoff}, \& {Babcock}}]{Gulbis2015}
{Gulbis}, A.~A.~S., {Emery}, J.~P., {Person}, M.~J., {et~al.} 2015, \icarus,
  246, 226

\bibitem[{{Hinson} {et~al.}(2017){Hinson}, {Linscott}, {Young}, {Tyler},
  {Stern}, {Beyer}, {Bird}, {Ennico}, {Gladstone}, {Olkin}, {P{\"a}tzold},
  {Schenk}, {Strobel}, {Summers}, {Weaver}, \& {Woods}}]{Hinson2017}
{Hinson}, D.~P., {Linscott}, I.~R., {Young}, L.~A., {et~al.} 2017, \icarus,
  290, 96

\bibitem[{{Holler} {et~al.}(2022){Holler}, {Yanez}, {Protopapa}, {Young},
  {Verbiscer}, {Chanover}, \& {Grundy}}]{Holler2022}
{Holler}, B.~J., {Yanez}, M.~D., {Protopapa}, S., {et~al.} 2022, \icarus, 373,
  114729

\bibitem[{{Hubbard} {et~al.}(1988){Hubbard}, {Hunten}, {Dieters}, {Hill}, \&
  {Watson}}]{Hubbard1988}
{Hubbard}, W.~B., {Hunten}, D.~M., {Dieters}, S.~W., {Hill}, K.~M., \&
  {Watson}, R.~D. 1988, \nat, 336, 452

\bibitem[{{Jacobson} {et~al.}(2019){Jacobson}, {Brozovic}, {Showalter},
  {Verbiscer}, {Buie}, \& {Helfenstein}}]{Jacobson2019}
{Jacobson}, R.~A., {Brozovic}, M., {Showalter}, M., {et~al.} 2019, in Pluto
  System After New Horizons, Vol. 2133, 7031

\bibitem[{{Lellouch} {et~al.}(2022){Lellouch}, {Butler}, {Moreno}, {Gurwell},
  {Lavvas}, {Bertrand}, {Fouchet}, {Strobel}, \& {Moullet}}]{Lellouch2022}
{Lellouch}, E., {Butler}, B., {Moreno}, R., {et~al.} 2022, \icarus, 372, 114722

\bibitem[{{Meza} {et~al.}(2019){Meza}, {Sicardy}, {Assafin}, {Ortiz},
  {Bertrand}, {Lellouch}, {Desmars}, {Forget}, {B{\'e}rard}, {Doressoundiram},
  {Lecacheux}, {Marques Oliveira}, {Roques}, {Widemann}, {Colas}, {Vachier},
  {Renner}, {Leiva}, {Braga-Ribas}, {Benedetti-Rossi}, {Camargo},
  {Dias-Oliveira}, {Morgado}, {Gomes-J{\'u}nior}, {Vieira-Martins}, {Behrend},
  {Tirado}, {Duffard}, {Morales}, {Santos-Sanz}, {Jel{\'\i}nek}, {Cunniffe},
  {Querel}, {Harnisch}, {Jansen}, {Pennell}, {Todd}, {Ivanov}, {Opitom},
  {Gillon}, {Jehin}, {Manfroid}, {Pollock}, {Reichart}, {Haislip}, {Ivarsen},
  {LaCluyze}, {Maury}, {Gil-Hutton}, {Dhillon}, {Littlefair}, {Marsh},
  {Veillet}, {Bath}, {Beisker}, {Bode}, {Kretlow}, {Herald}, {Gault}, {Kerr},
  {Pavlov}, {Farag{\'o}}, {Kl{\"o}s}, {Frappa}, {Lavayssi{\`e}re}, {Cole},
  {Giles}, {Greenhill}, {Hill}, {Buie}, {Olkin}, {Young}, {Young}, {Wasserman},
  {Devog{\`e}le}, {French}, {Bianco}, {Marchis}, {Brosch}, {Kaspi},
  {Polishook}, {Manulis}, {Ait Moulay Larbi}, {Benkhaldoun}, {Daassou}, {El
  Azhari}, {Moulane}, {Broughton}, {Milner}, {Dobosz}, {Bolt}, {Lade},
  {Gilmore}, {Kilmartin}, {Allen}, {Graham}, {Loader}, {McKay}, {Talbot},
  {Parker}, {Abe}, {Bendjoya}, {Rivet}, {Vernet}, {Di Fabrizio}, {Lorenzi},
  {Magazz{\'u}}, {Molinari}, {Gazeas}, {Tzouganatos}, {Carbognani}, {Bonnoli},
  {Marchini}, {Leto}, {Sanchez}, {Mancini}, {Kattentidt}, {Dohrmann}, {Guhl},
  {Rothe}, {Walzel}, {Wortmann}, {Eberle}, {Hampf}, {Ohlert}, {Krannich},
  {Murawsky}, {G{\"a}hrken}, {Gloistein}, {Alonso}, {Rom{\'a}n}, {Communal},
  {Jabet}, {deVisscher}, {S{\'e}rot}, {Janik}, {Moravec}, {Machado}, {Selva},
  {Perell{\'o}}, {Rovira}, {Conti}, {Papini}, {Salvaggio}, {Noschese},
  {Tsamis}, {Tigani}, {Barroy}, {Irzyk}, {Neel}, {Godard}, {Lanoisel{\'e}e},
  {Sogorb}, {V{\'e}rilhac}, {Bretton}, {Signoret}, {Ciabattari}, {Naves},
  {Boutet}, {De Queiroz}, {Lindner}, {Lindner}, {Enskonatus}, {Dangl},
  {Tordai}, {Eichler}, {Hattenbach}, {Peterson}, {Molnar}, \&
  {Howell}}]{Meza2019}
{Meza}, E., {Sicardy}, B., {Assafin}, M., {et~al.} 2019, \aap, 625, A42

\bibitem[{{Morgado} {et~al.}(2022){Morgado}, {Gomes-J{\'u}nior}, {Braga-Ribas},
  {Vieira-Martins}, {Desmars}, {Lainey}, {D'aversa}, {Dunham}, {Moore},
  {Bailli{\'e}}, {Herald}, {Assafin}, {Sicardy}, {Aoki}, {Bardecker}, {Barton},
  {Blank}, {Bruns}, {Carlson}, {Carlson}, {Cobble}, {Dunham}, {Eisfeldt},
  {Emilio}, {Jacques}, {Hinse}, {Kim}, {Malacarne}, {Maley}, {Maury}, {Meza},
  {Oliva}, {Orton}, {Pereira}, {Person}, {Plainaki}, {Sfair}, {Sindoni},
  {Smith}, {Sussenbach}, {Stuart}, {Vrolijk}, \& {Winter}}]{Morgado2022}
{Morgado}, B.~E., {Gomes-J{\'u}nior}, A.~R., {Braga-Ribas}, F., {et~al.} 2022,
  \aj, 163, 240

\bibitem[{{Morgado} {et~al.}(2021){Morgado}, {Sicardy}, {Braga-Ribas},
  {Desmars}, {Gomes-J{\'u}nior}, {B{\'e}rard}, {Leiva}, {Ortiz},
  {Vieira-Martins}, {Benedetti-Rossi}, {Santos-Sanz}, {Camargo}, {Duffard},
  {Rommel}, {Assafin}, {Boufleur}, {Colas}, {Kretlow}, {Beisker}, {Sfair},
  {Snodgrass}, {Morales}, {Fern{\'a}ndez-Valenzuela}, {Amaral}, {Amarante},
  {Artola}, {Backes}, {Bath}, {Bouley}, {Buie}, {Cacella}, {Colazo}, {Colque},
  {Dauvergne}, {Dominik}, {Emilio}, {Erickson}, {Evans}, {Fabrega-Polleri},
  {Garcia-Lambas}, {Giacchini}, {Hanna}, {Herald}, {Hesler}, {Hinse},
  {Jacques}, {Jehin}, {J{\o}rgensen}, {Kerr}, {Kouprianov}, {Levine}, {Linder},
  {Maley}, {Machado}, {Maquet}, {Maury}, {Melia}, {Meza}, {Mondon}, {Moura},
  {Newman}, {Payet}, {Pereira}, {Pollock}, {Poltronieri}, {Quispe-Huaynasi},
  {Reichart}, {de Santana}, {Schneiter}, {Sieyra}, {Skottfelt}, {Soulier},
  {Starck}, {Thierry}, {Torres}, {Trabuco}, {Unda-Sanzana}, {Yamashita},
  {Winter}, {Zapata}, \& {Zuluaga}}]{Morgado2021}
{Morgado}, B.~E., {Sicardy}, B., {Braga-Ribas}, F., {et~al.} 2021, \aap, 652,
  A141

\bibitem[{Newell \& Tiesinga(2019)}]{Newell2019}
Newell, D. \& Tiesinga, E. 2019, in The International System of Units (SI)

\bibitem[{{Olkin} {et~al.}(2015){Olkin}, {Young}, {Borncamp}, {Pickles},
  {Sicardy}, {Assafin}, {Bianco}, {Buie}, {de Oliveira}, {Gillon}, {French},
  {Ramos Gomes}, {Jehin}, {Morales}, {Opitom}, {Ortiz}, {Maury}, {Norbury},
  {Braga-Ribas}, {Smith}, {Wasserman}, {Young}, {Zacharias}, \&
  {Zacharias}}]{Olkin2015}
{Olkin}, C.~B., {Young}, L.~A., {Borncamp}, D., {et~al.} 2015, \icarus, 246,
  220

\bibitem[{{Park} {et~al.}(2021){Park}, {Folkner}, {Williams}, \&
  {Boggs}}]{Park2021}
{Park}, R.~S., {Folkner}, W.~M., {Williams}, J.~G., \& {Boggs}, D.~H. 2021,
  \aj, 161, 105

\bibitem[{{Pasachoff} {et~al.}(2017){Pasachoff}, {Babcock}, {Durst}, {Seeger},
  {Levine}, {Bosh}, {Person}, {Sickafoose}, {Zuluaga}, {Kosiarek}, {Abe},
  {Nagakane}, {Suzuki}, {Tristram}, \& {Arredondo}}]{Pasachoff2017}
{Pasachoff}, J.~M., {Babcock}, B.~A., {Durst}, R.~F., {et~al.} 2017, \icarus,
  296, 305

\bibitem[{{Pasachoff} {et~al.}(2005){Pasachoff}, {Souza}, {Babcock},
  {Ticehurst}, {Elliot}, {Person}, {Clancy}, {Roberts}, {Hall}, \&
  {Tholen}}]{Pasachoff2005}
{Pasachoff}, J.~M., {Souza}, S.~P., {Babcock}, B.~A., {et~al.} 2005, \aj, 129,
  1718

\bibitem[{{Pavlov}(2020)}]{Pavlov2020}
{Pavlov}, H. 2020, {Tangra: Software for video photometry and astrometry},
  Astrophysics Source Code Library, record ascl:2004.002

\bibitem[{{Pereira} {et~al.}(2023){Pereira}, {Sicardy}, {Morgado},
  {Braga-Ribas}, {Fern{\'a}ndez-Valenzuela}, {Souami}, {Holler}, {Boufleur},
  {Margoti}, {Assafin}, {Ortiz}, {Santos-Sanz}, {Epinat}, {Kervella},
  {Desmars}, {Vieira-Martins}, {Kilic}, {Gomes J{\'u}nior}, {Camargo},
  {Emilio}, {Vara-Lubiano}, {Kretlow}, {Albert}, {Alcock}, {Ball}, {Bender},
  {Buie}, {Butterfield}, {Camarca}, {Castro-Chac{\'o}n}, {Dunford}, {Fisher},
  {Gamble}, {Geary}, {Gnilka}, {Green}, {Hartman}, {Huang}, {Januszewski},
  {Johnston}, {Kagitani}, {Kamin}, {Kavelaars}, {Keller}, {de Kleer}, {Lehner},
  {Luken}, {Marchis}, {Marlin}, {McGregor}, {Nikitin}, {Nolthenius}, {Patrick},
  {Redfield}, {Rengstorf}, {Reyes-Ruiz}, {Seccull}, {Skrutskie}, {Smith},
  {Sproul}, {Stephens}, {Szentgyorgyi}, {S{\'a}nchez-Sanju{\'a}n}, {Tatsumi},
  {Verbiscer}, {Wang}, {Yoshida}, {Young}, \& {Zhang}}]{Pereira2023}
{Pereira}, C.~L., {Sicardy}, B., {Morgado}, B.~E., {et~al.} 2023, \aap, 673, L4

\bibitem[{{Person} {et~al.}(2021){Person}, {Bosh}, {Zuluaga}, {Sickafoose},
  {Levine}, {Pasachoff}, {Babcock}, {Dunham}, {McLean}, {Wolf}, {Abe},
  {Becklin}, {Bida}, {Bright}, {Brothers}, {Christie}, {Durst}, {Gilmore},
  {Hamilton}, {Harris}, {Johnson}, {Kilmartin}, {Kosiarek}, {Leppik},
  {Logsdon}, {Lucas}, {Mathers}, {Morley}, {Nelson}, {Ngan}, {Pf{\"u}ller},
  {Natusch}, {Sallum}, {Savage}, {Seeger}, {Siu}, {Stockdale}, {Suzuki},
  {Thanathibodee}, {Tilleman}, {Tristram}, {Vacca}, {Van Cleve}, {Varughese},
  {Weisenbach}, {Widen}, \& {Wiedemann}}]{Person2021}
{Person}, M.~J., {Bosh}, A.~S., {Zuluaga}, C.~A., {et~al.} 2021, \icarus, 356,
  113572

\bibitem[{{Person} {et~al.}(2013){Person}, {Dunham}, {Bosh}, {Levine},
  {Gulbis}, {Zangari}, {Zuluaga}, {Pasachoff}, {Babcock}, {Pandey}, {Amrhein},
  {Sallum}, {Tholen}, {Collins}, {Bida}, {Taylor}, {Bright}, {Wolf}, {Meyer},
  {Pfueller}, {Wiedemann}, {Roeser}, {Lucas}, {Kakkala}, {Ciotti}, {Plunkett},
  {Hiraoka}, {Best}, {Pilger}, {Micheli}, {Springmann}, {Hicks}, {Thackeray},
  {Emery}, {Tilleman}, {Harris}, {Sheppard}, {Rapoport}, {Ritchie}, {Pearson},
  {Mattingly}, {Brimacombe}, {Gault}, {Jones}, {Nolthenius}, {Broughton}, \&
  {Barry}}]{Person2013}
{Person}, M.~J., {Dunham}, E.~W., {Bosh}, A.~S., {et~al.} 2013, \aj, 146, 83

\bibitem[{{Press} {et~al.}(2007){Press}, {Teukolsky}, {Vetterling}, \&
  {Flannery}}]{Press2007}
{Press}, W.~H., {Teukolsky}, S.~A., {Vetterling}, W.~T., \& {Flannery}, B.~P.
  2007, {Numerical recipes in C++ : the art of scientific computing (3rd
  Edition)}, 3rd edn. (Cambridge University Press)

\bibitem[{{Rannou} \& {Durry}(2009)}]{Rannou2009}
{Rannou}, P. \& {Durry}, G. 2009, Journal of Geophysical Research (Planets),
  114, E11013

\bibitem[{{Sicardy}(2022)}]{Sicardy2022}
{Sicardy}, B. 2022, arXiv e-prints, arXiv:2206.06236

\bibitem[{{Sicardy} {et~al.}(2021){Sicardy}, {Ashok}, {Tej}, {Pawar},
  {Deshmukh}, {Deshpande}, {Sharma}, {Desmars}, {Assafin}, {Ortiz},
  {Benedetti-Rossi}, {Braga-Ribas}, {Vieira-Martins}, {Santos-Sanz}, {Chand},
  \& {Bhatt}}]{Sicardy2021}
{Sicardy}, B., {Ashok}, N.~M., {Tej}, A., {et~al.} 2021, \apjl, 923, L31

\bibitem[{{Sicardy} {et~al.}(2011){Sicardy}, {Bolt}, {Broughton}, {Dobosz},
  {Gault}, {Kerr}, {B{\'e}nard}, {Frappa}, {Lecacheux}, {Peyrot},
  {Teng-Chuen-Yu}, {Beisker}, {Boissel}, {Buckley}, {Colas}, {de Witt},
  {Doressoundiram}, {Roques}, {Widemann}, {Gruhn}, {Batista}, {Biggs},
  {Dieters}, {Greenhill}, {Groom}, {Herald}, {Lade}, {Mathers}, {Assafin},
  {Camargo}, {Vieira-Martins}, {Andrei}, {da Silva Neto}, {Braga-Ribas}, \&
  {Behrend}}]{Sicardy2011a}
{Sicardy}, B., {Bolt}, G., {Broughton}, J., {et~al.} 2011, \aj, 141, 67

\bibitem[{{Sicardy} {et~al.}(2016){Sicardy}, {Talbot}, {Meza}, {Camargo},
  {Desmars}, {Gault}, {Herald}, {Kerr}, {Pavlov}, {Braga-Ribas}, {Assafin},
  {Benedetti-Rossi}, {Dias-Oliveira}, {Gomes-J{\'u}nior}, {Vieira-Martins},
  {B{\'e}rard}, {Kervella}, {Lecacheux}, {Lellouch}, {Beisker}, {Dunham},
  {Jel{\'\i}nek}, {Duffard}, {Ortiz}, {Castro-Tirado}, {Cunniffe}, {Querel},
  {Yock}, {Cole}, {Giles}, {Hill}, {Beaulieu}, {Harnisch}, {Jansen}, {Pennell},
  {Todd}, {Allen}, {Graham}, {Loader}, {McKay}, {Milner}, {Parker}, {Barry},
  {Bradshaw}, {Broughton}, {Davis}, {Devillepoix}, {Drummond}, {Field},
  {Forbes}, {Giles}, {Glassey}, {Groom}, {Hooper}, {Horvat}, {Hudson},
  {Idaczyk}, {Jenke}, {Lade}, {Newman}, {Nosworthy}, {Purcell}, {Skilton},
  {Streamer}, {Unwin}, {Watanabe}, {White}, \& {Watson}}]{Sicardy2016}
{Sicardy}, B., {Talbot}, J., {Meza}, E., {et~al.} 2016, \apjl, 819, L38

\bibitem[{{Sicardy} {et~al.}(2003){Sicardy}, {Widemann}, {Lellouch}, {Veillet},
  {Cuillandre}, {Colas}, {Roques}, {Beisker}, {Kretlow}, {Lagrange}, {Gendron},
  {Lacombe}, {Lecacheux}, {Birnbaum}, {Fienga}, {Leyrat}, {Maury}, {Raynaud},
  {Renner}, {Schultheis}, {Brooks}, {Delsanti}, {Hainaut}, {Gilmozzi},
  {Lidman}, {Spyromilio}, {Rapaport}, {Rosenzweig}, {Naranjo}, {Porras},
  {D{\'\i}az}, {Calder{\'o}n}, {Carrillo}, {Carvajal}, {Recalde}, {Cavero},
  {Montalvo}, {Barr{\'\i}a}, {Campos}, {Duffard}, \& {Levato}}]{Sicardy2003}
{Sicardy}, B., {Widemann}, T., {Lellouch}, E., {et~al.} 2003, \nat, 424, 168

\bibitem[{{Silva-Cabrera} {et~al.}(2022){Silva-Cabrera}, {Castro-Chac{\'o}n},
  {Reyes-Ruiz}, {Lehner}, {Guerrero}, {Huang}, {Alvarez-Santana}, {Chang},
  {Ling}, {Porras-Navarro}, {Hern{\'a}ndez-{\'A}guila}, {P{\'e}rez-Arce},
  {Rojas-Quintero}, {Avila}, {Wang}, {Alcock}, {Chen}, {Granados Contreras},
  {Cook}, {Geary}, {Hern{\'a}ndez-Valencia}, {Kavelaars}, {Norton},
  {S{\'a}nchez}, {Szentgyorgyi}, {Yen}, \& {Zhang}}]{SilvaCabrera2022}
{Silva-Cabrera}, J.~S., {Castro-Chac{\'o}n}, J.~H., {Reyes-Ruiz}, M., {et~al.}
  2022, \mnras, 511, 5550

\bibitem[{{Stern} {et~al.}(2015){Stern}, {Bagenal}, {Ennico}, {Gladstone},
  {Grundy}, {McKinnon}, {Moore}, {Olkin}, {Spencer}, {Weaver}, {Young},
  {Andert}, {Andrews}, {Banks}, {Bauer}, {Bauman}, {Barnouin}, {Bedini},
  {Beisser}, {Beyer}, {Bhaskaran}, {Binzel}, {Birath}, {Bird}, {Bogan},
  {Bowman}, {Bray}, {Brozovic}, {Bryan}, {Buckley}, {Buie}, {Buratti},
  {Bushman}, {Calloway}, {Carcich}, {Cheng}, {Conard}, {Conrad}, {Cook},
  {Cruikshank}, {Custodio}, {Dalle Ore}, {Deboy}, {Dischner}, {Dumont},
  {Earle}, {Elliott}, {Ercol}, {Ernst}, {Finley}, {Flanigan}, {Fountain},
  {Freeze}, {Greathouse}, {Green}, {Guo}, {Hahn}, {Hamilton}, {Hamilton},
  {Hanley}, {Harch}, {Hart}, {Hersman}, {Hill}, {Hill}, {Hinson}, {Holdridge},
  {Horanyi}, {Howard}, {Howett}, {Jackman}, {Jacobson}, {Jennings}, {Kammer},
  {Kang}, {Kaufmann}, {Kollmann}, {Krimigis}, {Kusnierkiewicz}, {Lauer}, {Lee},
  {Lindstrom}, {Linscott}, {Lisse}, {Lunsford}, {Mallder}, {Martin}, {McComas},
  {McNutt}, {Mehoke}, {Mehoke}, {Melin}, {Mutchler}, {Nelson}, {Nimmo},
  {Nunez}, {Ocampo}, {Owen}, {Paetzold}, {Page}, {Parker}, {Parker},
  {Pelletier}, {Peterson}, {Pinkine}, {Piquette}, {Porter}, {Protopapa},
  {Redfern}, {Reitsema}, {Reuter}, {Roberts}, {Robbins}, {Rogers}, {Rose},
  {Runyon}, {Retherford}, {Ryschkewitsch}, {Schenk}, {Schindhelm}, {Sepan},
  {Showalter}, {Singer}, {Soluri}, {Stanbridge}, {Steffl}, {Strobel}, {Stryk},
  {Summers}, {Szalay}, {Tapley}, {Taylor}, {Taylor}, {Throop}, {Tsang},
  {Tyler}, {Umurhan}, {Verbiscer}, {Versteeg}, {Vincent}, {Webbert}, {Weidner},
  {Weigle}, {White}, {Whittenburg}, {Williams}, {Williams}, {Williams},
  {Woods}, {Zangari}, \& {Zirnstein}}]{Stern2015}
{Stern}, S.~A., {Bagenal}, F., {Ennico}, K., {et~al.} 2015, Science, 350,
  aad1815

\bibitem[{{Washburn} {et~al.}(1930){Washburn}, {West}, {Dorsey}, (U.S.),
  Council, \& of~Sciences~(U.S.)}]{Washburn1930}
{Washburn}, E.~W., {West}, C.~J., {Dorsey}, N.~E., {et~al.} 1930, International
  Critical Tables of Numerical Data, Physics, Chemistry and Technology
  (Washington, DC: The National Academies Press)

\bibitem[{{Yelle} \& {Elliot}(1997)}]{Yelle1997}
{Yelle}, R.~V. \& {Elliot}, J.~L. 1997, in Pluto and Charon, ed. S.~A. {Stern}
  \& D.~J. {Tholen}, 347

\bibitem[{{Young} {et~al.}(2021){Young}, {Young}, {Johnson}, \& {PHOT
  Team}}]{Young2021}
{Young}, E., {Young}, L.~A., {Johnson}, P.~E., \& {PHOT Team}. 2021, in
  AAS/Division for Planetary Sciences Meeting Abstracts, Vol.~53, AAS/Division
  for Planetary Sciences Meeting Abstracts, 307.06

\bibitem[{{Young} {et~al.}(2008){Young}, {French}, {Young}, {Ruhland}, {Buie},
  {Olkin}, {Regester}, {Shoemaker}, {Blow}, {Broughton}, {Christie}, {Gault},
  {Lade}, \& {Natusch}}]{Young2008}
{Young}, E.~F., {French}, R.~G., {Young}, L.~A., {et~al.} 2008, \aj, 136, 1757

\end{thebibliography}

\begin{appendix}
   
   \section{Two occultation campaigns in 2019}
   \label{app:cam}
   

   {The 17 July 2019\footnote{\url{https://lesia.obspm.fr/lucky-star/occ.php?p=13163}} and the 5 September 2019\footnote{\url{https://lesia.obspm.fr/lucky-star/occ.php?p=13166}} stellar occultations by Pluto were originally predicted by the ERC Lucky Star project\footnote{\url{https://lesia.obspm.fr/lucky-star}}.}
   
   \subsection{The 17 July 2019 occultation campaign}
 
\begin{figure*}[h]
      \centering
      \begin{subfigure}{0.46\linewidth}
         \includegraphics[width=\linewidth]{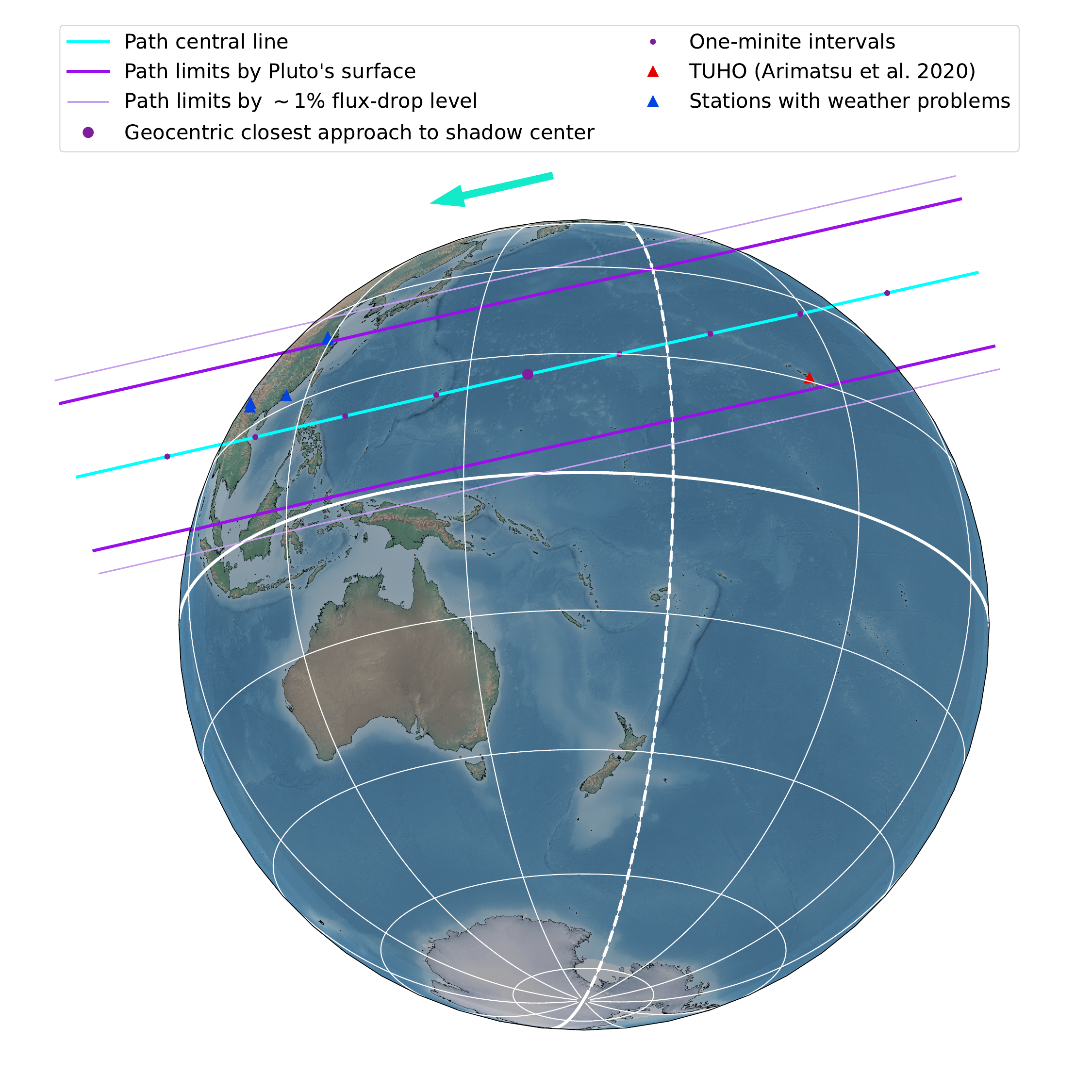}
         \caption{}
         \label{fig:occmap201907:a}
      \end{subfigure}
      \hfil
      \begin{subfigure}{0.46\linewidth}
         \includegraphics[width=\linewidth]{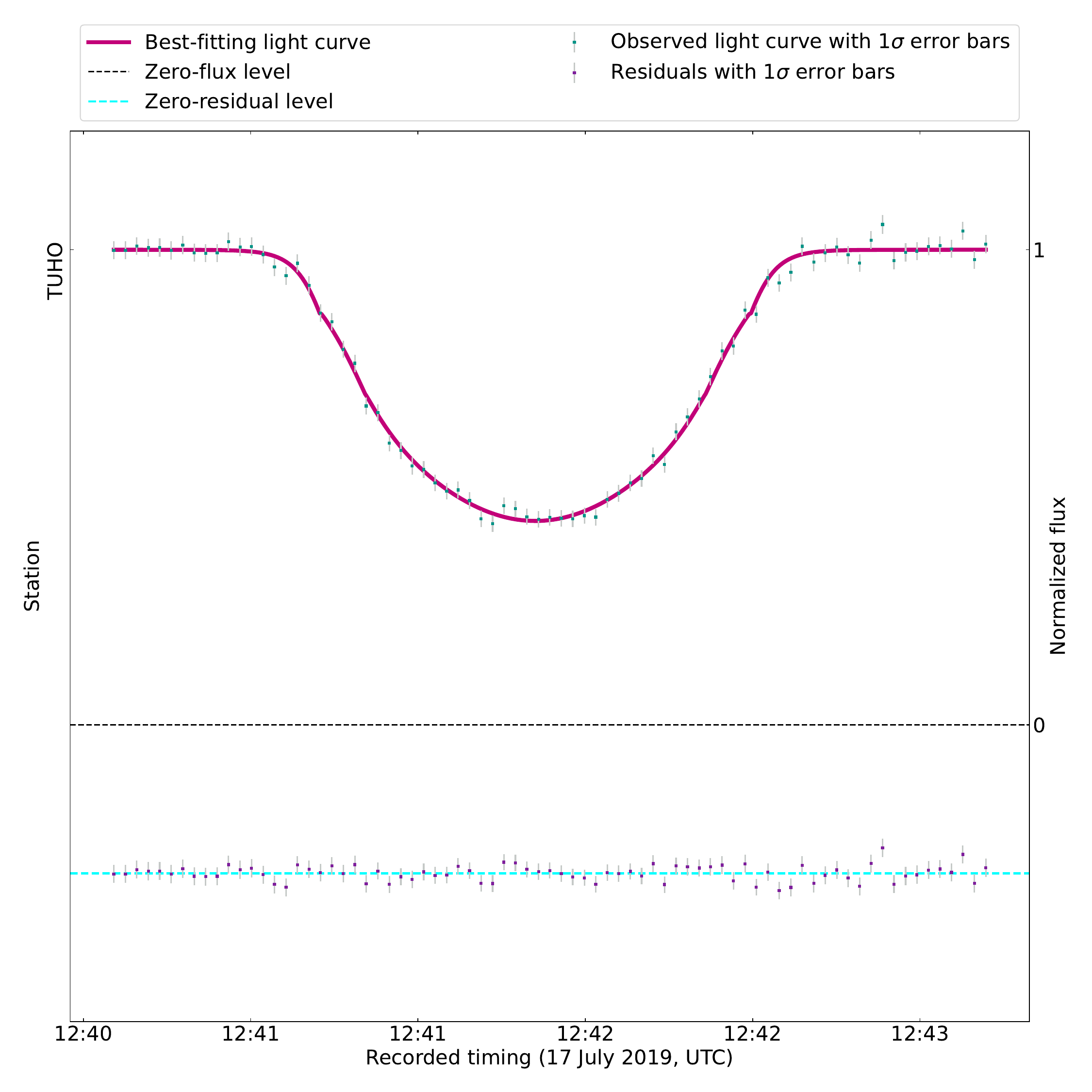}
         \caption{}
         \label{fig:occmap201907:b}
      \end{subfigure}

      \smallskip
      
      \begin{subfigure}{0.46\linewidth}
         \includegraphics[width=\linewidth]{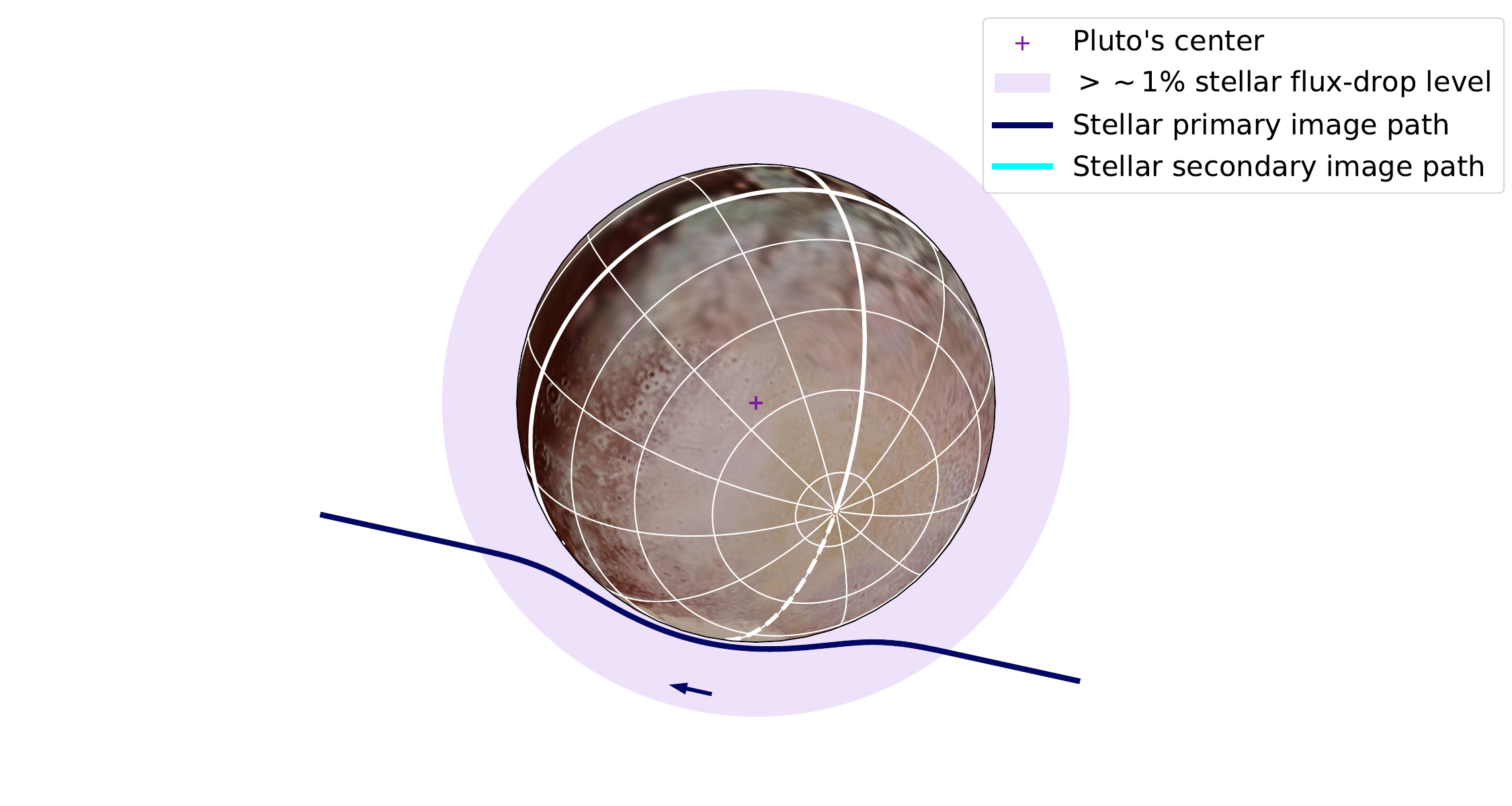}
         \caption{}
         \label{fig:occmap201907:c}
      \end{subfigure}
      \hfil
      \begin{subfigure}{0.46\linewidth}
         \includegraphics[width=\linewidth]{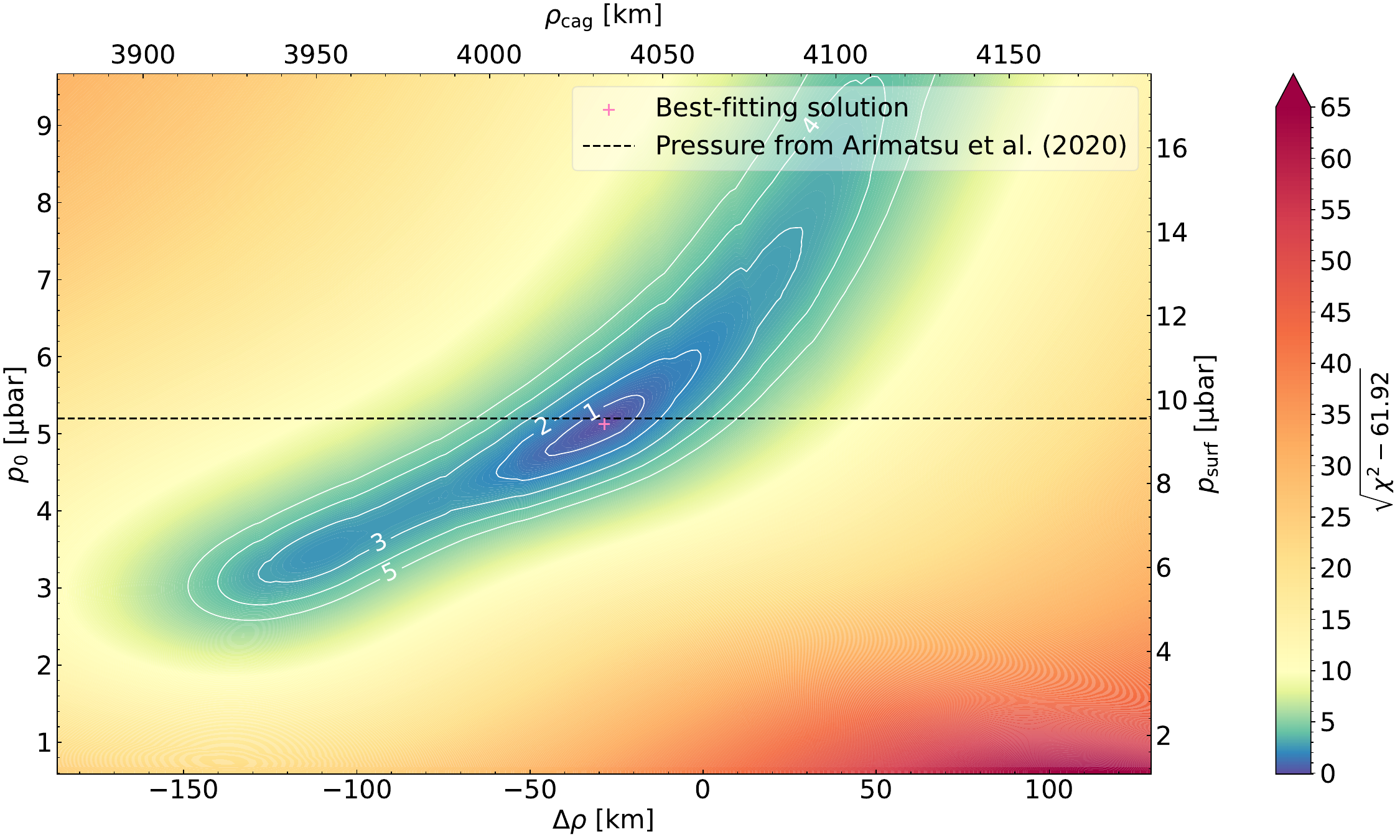}
         \caption{}
         \label{fig:occmap201907:d}
      \end{subfigure} 
      \caption{Reanalysis of the 17 July 2019 event based on the TUHO {observational data from Figure 1 of \citet{Arimatsu2020}} {(credit: Arimatsu et al, A\&A, 638, L5, 2020, reproduced with permission \copyright~ESO)}. 
      Panel (a): Reconstructed occultation map. 
      Panel (b): Observed and simultaneously fitted light curves.
      Panel (c): Reconstructed stellar paths seen by TUHO.
      Panel (d): The $\chi^2$ map, where $\chi^2$ denotes the goodness-of-fit value using these data. 
      The best-fitting $\chi^2$ value per degree of freedom is $0.860$.}
      \label{fig:occmap201907}
   \end{figure*}

   Figure \ref{fig:occmap201907:a} presents the reconstructed path of the shadow of Pluto\footnote{The occulted star is Gaia DR3 \href{https://vizier.cds.unistra.fr/viz-bin/VizieR-S?Gaia\%20DR3\%206772059623498733952}{$6772059623498733952$}, of which the astrometric and photometric parameters are obtained from VizieR.} during the 17 July 2019 event, which was studied by \citet{Arimatsu2020}.
   We also organized an observation campaign involving stations detailed in Table \ref{tab:err201907}. 
   All the above-mentioned stations are also presented in Figure \ref{fig:occmap201907:a}.
   Unfortunately, all our stations encountered weather problems.

   \begin{table*}[h]
      \centering
      \caption{Stations that encountered weather problems on 17 July 2019}\label{tab:err201907}
      \begin{tabular}{lrrrl}
         \hline\hline
         Station & Longitude (N) & Latitude (E) & Altitude (m) & Observers \\
         \hline
         PMOXL  &  $118\degr57'30\farcs0$  &  $32\degr07'33\farcs0$  &  $74$  &  Jian Chen, Ansheng Zhu, Yue Chen  \\ 
         XYOS  &  $118\degr27'50\farcs0$  &  $32\degr44'03\farcs7$  &  $227$  &  Wei Zhang  \\ 
         YAOS  &  $101\degr10'52\farcs0$  &  $25\degr31'43\farcs0$  &  $1943$  &  Ye Yuan, Fan Li  \\ 
         GMG  &  $100\degr01'51\farcs6$  &  $26\degr42'33\farcs1$  &  $3185$  &  Xiaoli Wang, Jianguo Wang  \\ 
         SZAO  &  $114\degr33'21\farcs0$  &  $22\degr28'56\farcs0$  &  $220$  &  Jiahui Ye, Yijing Zhu, Delai Li, Lin Mei  \\ 
         \hline
      \end{tabular}
   \end{table*}

   \subsection{The 5 September 2019 occultation campaign}

   The occultation map of the 5 September 2019 event is given by Figure \ref{fig:occmap201909}.
   Table \ref{tab:pos201909} lists the circumstances of stations with positive detections.
   Table \ref{tab:err201909} lists the circumstances of stations that encountered weather problems.

   \begin{table*}[h]
      \centering
      \caption{Stations with positive detections on 5 September 2019}\label{tab:pos201909}
      \begin{tabular}{lrcrl}
         \hline\hline
         Station & Longitude (N) & Telescope & Exposure (s) & Observers \\
         ~       & Latitude (E)  & Camera    & Circle (s)   &  \\
         ~       & Altitude (m)  & Filter    &              &  \\
         \hline
         DWM   & $114\degr06'44\farcs5$ & DOB 0.4 m                                &                  0.2 & Ye Yuan, Jian Chen, \\
         ~     & $28\degr25'28\farcs0$ & QHY174GPS                                &                  0.2 &  Wei Tan, Ying Wang, \\
         ~     & $1388$ & clear                                    &  &   Hui Zhang, Shun Yao \\
         \hline
         HNU   & $114\degr30'58\farcs0$ & RC 0.5 m                                 &                    9 & Qiang Zhang, \\
               & $37\degr59'55\farcs0$ & CCD                                      &                   14 &  Fan Li, \\
               & $69$ & clear                                    &                        & Zhensen Fu \\
         \hline
      \end{tabular}
   \end{table*}

   \begin{table*}[h]
      \centering
      \caption{Stations that encountered weather problems on 5 September 2019}\label{tab:err201909}
      \begin{tabular}{lrrrl}
         \hline\hline
         Station & Longitude (N) & Latitude (E) & Altitude (m) & Observers \\
         \hline
         PMOXL  &  $118\degr57'30\farcs0$  &  $32\degr07'33\farcs0$  &  $74$  &  Jian Chen, Ansheng Zhu, Yue Chen  \\ 
         NJXD  &  $118\degr25'17\farcs5$  &  $32\degr04'40\farcs8$  &  $74$  &  Jun Xu  \\ 
         XYOS  &  $118\degr27'50\farcs0$  &  $32\degr44'03\farcs7$  &  $227$  &  Wei Zhang  \\ 
         YAOS  &  $101\degr10'52\farcs0$  &  $25\degr31'43\farcs0$  &  $1943$  &  Chen Zhang, Fan Li, Ye Yuan  \\ 
         GMG  &  $100\degr01'51\farcs6$  &  $26\degr42'33\farcs1$  &  $3185$  &  Xiliang Zhang, Xiaoli Wang, Jianguo Wang  \\ 
         SZAO  &  $114\degr33'21\farcs0$  &  $22\degr28'56\farcs0$  &  $220$  &  Jiahui Ye, Delai Li, Lin Mei  \\ 
         QHOS  &  $97\degr33'36\farcs0$  &  $37\degr22'24\farcs0$  &  $3200$  &  Wei Zhang  \\ 
         QDAS  &  $119\degr56'38\farcs0$  &  $36\degr07'20\farcs0$  &  $107$  &  Kun Zhou  \\ 
         \hline
      \end{tabular}
   \end{table*}

   \section{Data processing}
   \label{app:data}


   The observational data were initially obtained in the \texttt{FITS} format, a commonly used format for astronomical data.
   To process and analyze these data, we used the \texttt{Tangra} occultation photometric tool developed by \citet{Pavlov2020}. 
   This tool offers various functionalities and algorithms specifically designed for analyzing occultation events.
   When calibration images including bias, dark, and flat-field frames are available, they are used for image correction.
   The default measurement type and tracking method provided by \texttt{Tangra} were applied.
   The reduction method was aperture photometry with median background subtraction.
   The combined image of the occulted star and Pluto's system is circled as the target.

   For each station, a selection of suitable guiding and nearby reference stars (typically three) was made.
   All images of target and reference stars are confirmed as unaffected by overexposure.  
   The signal flux counts, denoted as $I_\txt{signal}$, and the background flux counts, denoted as $I_\txt{bkg}$, were measured for the target and each reference star in all \texttt{FITS} files. 
   The timing information and measurement results were exported to a \texttt{CSV} file specific to each station, from which the observed light curves would be generated.

   Further data processing and analysis were performed using our data reduction code implemented in \texttt{Python}.
   Each flux measurement for the target or a reference star was calculated as $I = I_\txt{signal} - I_\txt{bkg}$. 
   The measurement error, $\sigma_{I}$, was modeled using the classical form given by:
   \begin{equation}
      \sigma_I = \sqrt{\sigma_{\txt{bkg}}^2 + I \cdot g_{\txt{eff}}^{-1}},
   \end{equation}
   where 
   $\sigma_{\txt{bkg}}$ is the standard deviation of the background ($I_{\txt{bkg}}$) and
   $g_\txt{eff}$ is the effective gain that represents the number of electrons per flux count.

   The relative photometric result of the target is defined as
   \begin{equation}
      F = \frac{I_T}{I_{RS}},
   \end{equation}
   where 
   $I_T$ is the $I$ of target and $I_{RS}$ the sum of those of all reference stars. 
   This step helps eliminate low-frequency variations caused by atmospheric and instrumental effects. 
   The corresponding measurement error, $\sigma_{F}$, was derived using the usual formula for propagation of errors \citep{Press2007}.

   If the effective gain $g_\txt{eff}$ was not precisely known, $g_\txt{eff}^{-1}$ was adjusted with the non-negative constraint to fit the derived $\sigma_{F}$ model ($\sigma_{F}(g_\txt{eff}^{-1})$ where $g_\txt{eff}^{-1} \ge 0$) to the standard deviation of $F$ outside the occultation part.

   Finally, the normalized total observed flux, denoted as $f$, and its error, $\sigma$, were derived by dividing $F$ and $\sigma_{F}$, respectively, by the median of $F$ outside the occultation part. As a result, $f$ outside the occultation part approximates unity.

   \section{Synthetic light curve model for a stellar occultation by Pluto}
   \label{app:lcmod}


   For a given station, the light-curve model is represented as:
   \begin{equation}
      \phi (t) = A \cdot \left(s \cdot \psi (t) + (1-s )\right),
   \end{equation}
   where 
   $t$ is the recorded timing of the observation and $\phi$ is the received flux, as a function of $t$;
   $A$ is the total flux of the star and Pluto's system when the star is not occulted;
   $s$ is the flux ratio of the occulted star to $A$; and
   $\psi$ is the normalized (between zero and unity) flux of the occulted star, which represents the total flux of the primary (sometimes called the near-limb) and secondary (or far-limb, when available) images produced by Pluto's spherical planetary atmosphere \citep{Sicardy2022} as shown in Figure \ref{fig:obs201909}.

   To calculate $\psi$, the following equation is used:
   \begin{equation}
      \psi(t) = \left|\frac{r_+}{z}\right| \left|\frac{dr_+}{dz}\right| + \left|\frac{r_-}{z}\right| \left|\frac{dr_-}{dz}\right|, \quad (z > 0, r_{\pm} > R_\txt{p})
   \end{equation}
   where
   $z$ is the distance of the station to the shadow axis, that is, the line passing through the center of Pluto and parallel to the light ray from the star;
   $r_+$ and $r_-$ represent the closest approach distances of the primary and secondary images to the shadow axis, respectively; and
   $R_\txt{p}$ is the radius of Pluto's body.
   The dividers, $|r_\pm/z|$, and derivatives, $|dr_\pm/dz|$, account for the compression and stretching effects produced by the spherical planetary atmosphere, respectively, caused by limb curvature and differential refraction. 
 
   Given reference ephemerides and a star catalog, the distance $z$ is modeled in the International Celestial Reference Frame (ICRF) using the following equation:
   \begin{equation}
      z(t) = \left|\left(\bm{\rho}_\txt{p}(t+\Delta t_i+\Delta\tau) + \Delta\rho \cdot \hat{\bm{n}}_\txt{cag} - \bm{\rho}_i(t+\Delta t_i)\right) \bm{\times} \hat{\bm \ell}_\txt{s}\right|,
      \label{eq:z}
   \end{equation}
   where 
   $t+\Delta t_i$ represents the real observation epoch corrected for the camera time recording offset $\Delta t_i$; 
   $\hat{\bm \ell}_\txt{s}$ is the unit vector in the direction of the geocentric astrometric position of the occulted star; 
   $\bm{\rho}_i$ is the geocentric geometric position of $i$;
   $\bm{\rho}_\txt{p}$ is the geocentric astrometric position of Pluto obtained at $t+\Delta t_i+\Delta\tau$, with $\Delta\tau$ accounting for the ephemeris offset along the geocentric motion of Pluto;
   $\hat{\bm{n}}_\txt{cag}$ is the unit vector in the northern direction along the predicted geocentric closest approach to the shadow axis, multiplied by $\Delta\rho$ to account for the ephemeris offset across the geocentric motion of Pluto.
   The two ephemeris offset parameters, $\Delta\tau$ and $\Delta\rho$, represent the corrections to the epoch $t_\txt{cag}$ and distance $\rho_\txt{cag}$ of the geoncetric closest approach predicted with the given reference ephemerides and star catalog (see Section \ref{ssect:lcmod}).

   Then, the values of $r_\pm$ can be numerically solved using the relationship between $z$ and $r_\pm$:
   \begin{equation}
      z = \pm (r_\pm + D\cdot\omega(r_\pm)),
   \end{equation}
   where 
   $D$ represents the light travel distance, which can be approximated by the geocentric distance of Pluto, and $\omega(r)$ is the total deviation angle at $r$.

   As detailed in \citet{DiasOliveira2015} and \citet{Sicardy2022}, $\omega(r)$ can be obtained by a ray-tracing code that follows these steps:
   \begin{itemize}
      \item set the temperature profile $T(r)$ of Pluto using Equation (4) from \citet{DiasOliveira2015}, with parameters obtained from \citet{Sicardy2016};
      \item set the gas molecular refractivity $K$ corresponding to a stellar wavelength of $\lambda_\microm$;
      \item set the atmospheric pressure $p_0$ at the reference radius $r_0$ of 1215 km either directly, or determined from a surface pressure $p_{\txt{surf}}$ (at $R_\txt{p} = 1187$ km) with the ratio $p_{\txt{surf}}/p_0 = 1.837$ used by previous studies \citep{Meza2019,Sicardy2021}; 
      \item set the needed boundary condition for the gas molecular density, $n_0 = p_0/(k_\txt{B} \cdot T(r_0))$, where $k_\txt{B}$ is Boltzmann's constant; 
      \item {integrate} the first-order differential equation (Equations (5) and (6) of \citet{DiasOliveira2015}) to obtain the gas molecular density profile $n(r)$;
      \item transform $n(r)$ to the gas refractivity $\nu(r)$ using $\nu(r) = K \cdot n(r)$ (Equation (7) from \citet{DiasOliveira2015});
      \item calculate the total deviation based on the straight line approximation (Equation (16) of \citet{Sicardy2022}, with the upper limit of $r$ set as 2600 km).
    \end{itemize}
   Finally, an interpolator of $\omega(r)$ for specific $p_0$ and $K$ values is built in advance. 
   This allows efficient computation of $\omega(r)$ during the analysis of the occultation observations.

   \section{Results of the reanalyzed events necessary for comparisons}
   \label{app:reana}
   
   \subsection{The 15 August 2018 event}
   \label{sapp:2018}

\begin{figure*}[h]
      \centering
      \begin{subfigure}{0.46\linewidth}
         \includegraphics[width=\linewidth]{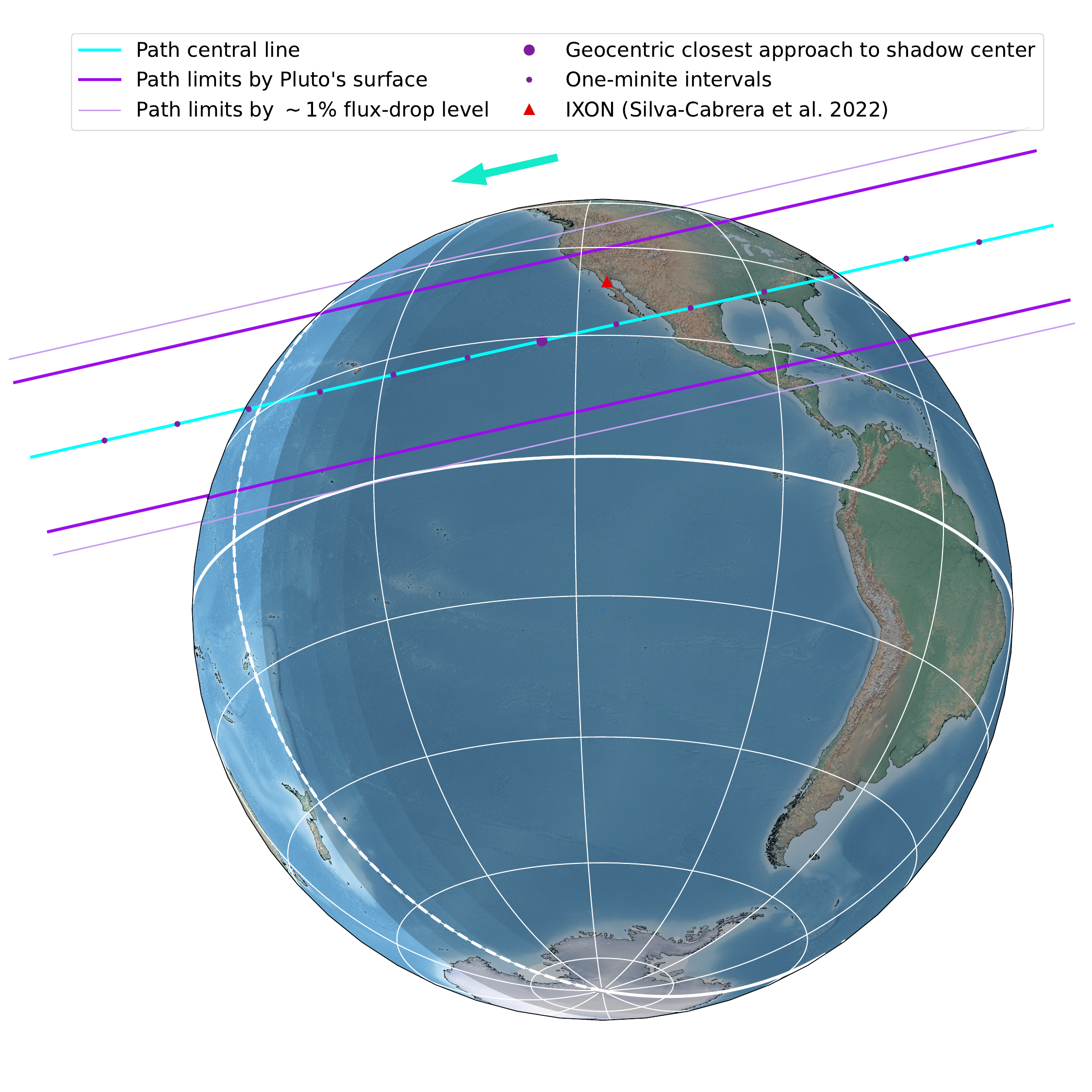}
         \caption{} 
         \label{fig:occmap201808:a}
      \end{subfigure}
      \hfil
      \begin{subfigure}{0.46\linewidth}
         \includegraphics[width=\linewidth]{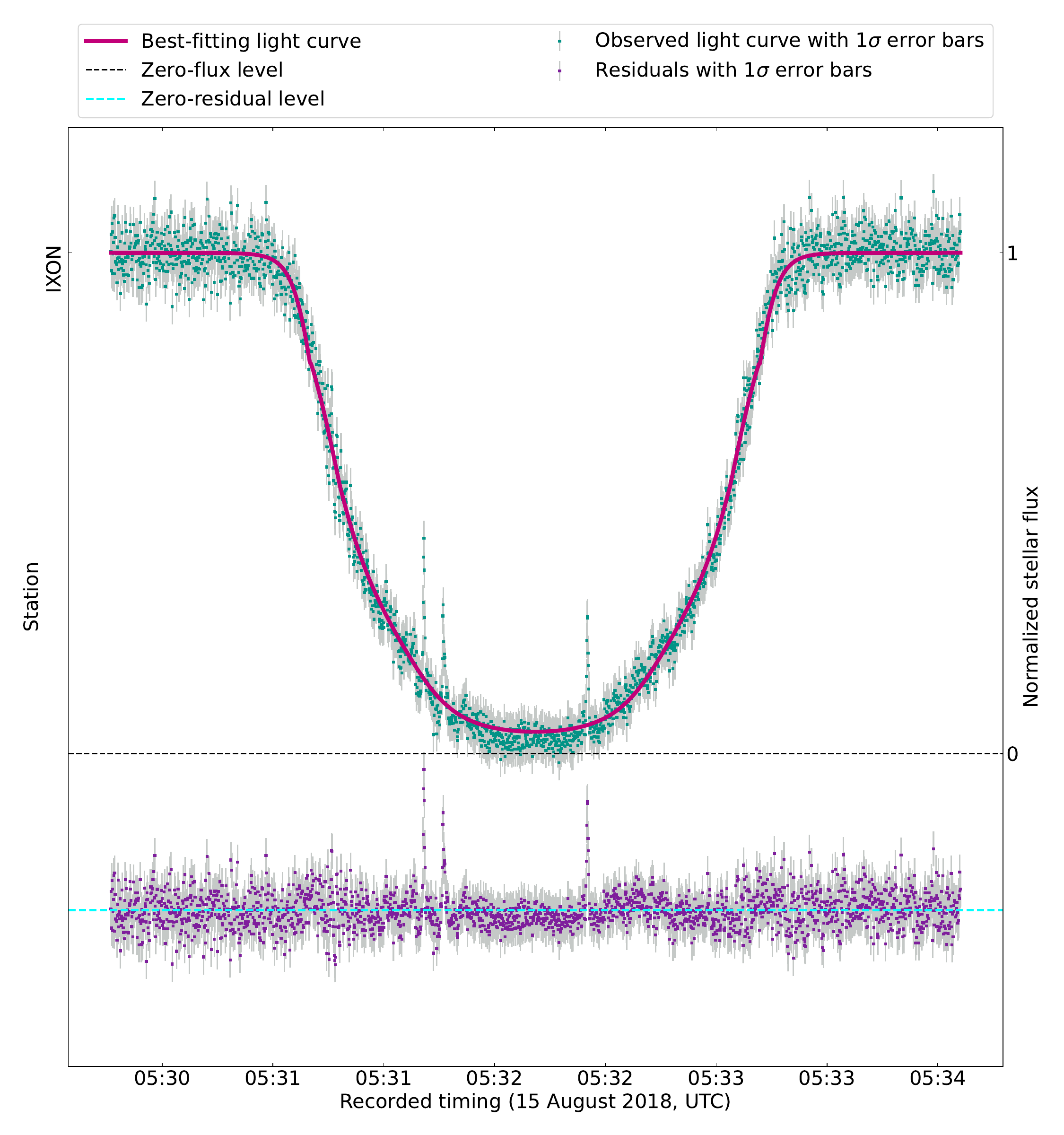}
         \caption{} 
         \label{fig:occmap201808:b}
      \end{subfigure}

      \smallskip
      
      \begin{subfigure}{0.46\linewidth}
         \includegraphics[width=\linewidth]{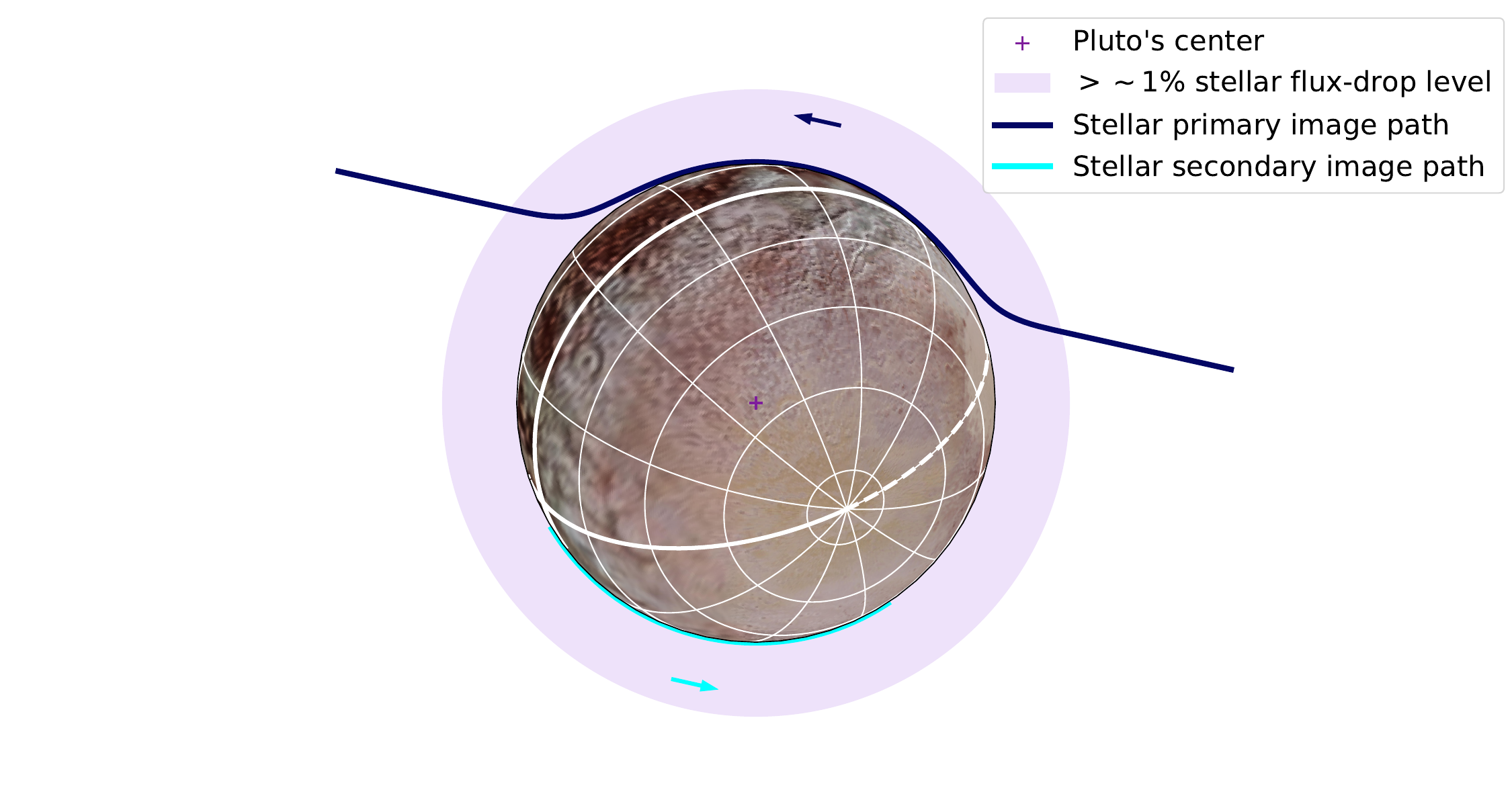}
         \caption{} 
         \label{fig:occmap201808:c}
      \end{subfigure}
      \hfil
      \begin{subfigure}{0.46\linewidth}
         \includegraphics[width=\linewidth]{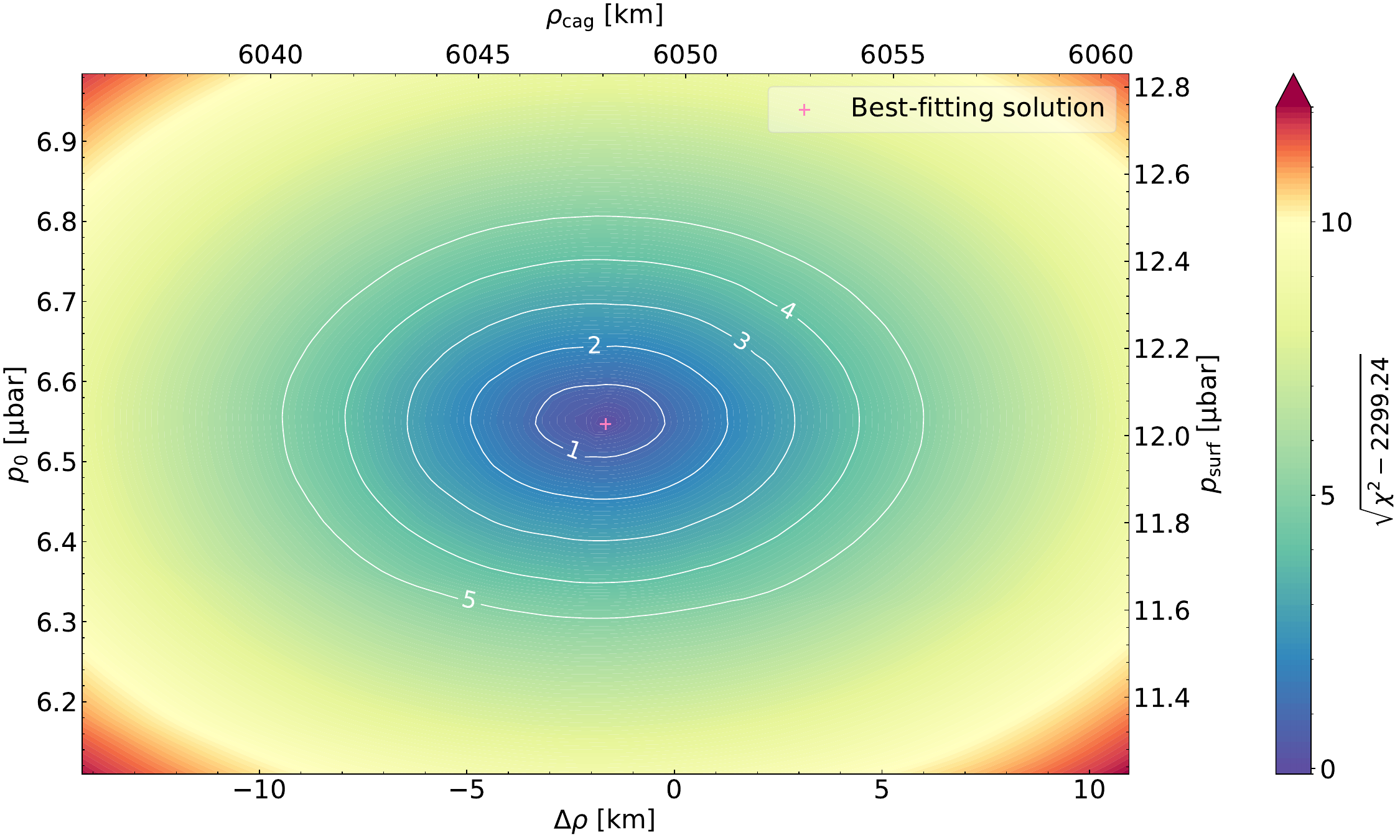}
         \caption{}
         \label{fig:occmap201808:d}
      \end{subfigure} 
      \caption{Reanalysis of the 15 August 2018 event based on the IXON {observational data from the online supplemental data file of \citet{SilvaCabrera2022}}. The raw data were normalized, with the flux contribution of the Pluto system carefully removed by \citet{SilvaCabrera2022}.
      The uncertainties used for weighting are the standard deviation of the data outside the occultation part for each station.
      Panel (a): Reconstructed occultation map. 
      Panel (b): Observed and simultaneously fitted light curves.
      Panel (c): Reconstructed stellar paths seen by IXON.
      Panel (d): The $\chi^2$ map, where $\chi^2$ denotes the goodness-of-fit value using these data. 
      The best-fitting $\chi^2$ value per degree of freedom is ${1.046}$.}
      \label{fig:occmap201808}
   \end{figure*}

   {Based on the IXON observational data from the online supplemental data file\footnote{{``stac401\_Supplement\_File - zip file'', via the paper link \url{https://doi.org/10.1093/mnras/stac401}}.} of} \citet{SilvaCabrera2022}, we reanalyzed the 15 August 2018 event\footnote{The occulted star is Gaia DR3 \href{https://vizier.cds.unistra.fr/viz-bin/VizieR-S?Gaia\%20DR3\%206772629170525258240}{$6772629170525258240$}, of which the astrometric and photometric parameters are obtained from {VizieR}.} using our fitting procedure.
   As the raw data were normalized, with the flux contribution of the Pluto system carefully removed by \citet{SilvaCabrera2022}, all $A_i$ and $s_i$ were fixed as unity in our fitting procedure.
   Moreover, as the uncertainties used for weighting are not given, we used the standard deviation of the data outside the occultation part for each station.
   Results are presented in Figure \ref{fig:occmap201808}.
   Our surface pressure remeasurement is ${12.027}_{-0.08}^{+0.09}~\microbar$.

   \subsection{The 17 July 2019 event}
   \label{sapp:2019}

   Based on the TUHO observational data {and uncertainties} {from Figure 1 of \citet{Arimatsu2020}} {(credit: Arimatsu et al, A\&A, 638, L5, 2020, reproduced with permission \copyright~ESO)}, {we reanalyzed} the 17 July 2019 event using our fitting procedure.
   Results are presented in Figure \ref{fig:occmap201907}.
   Our surface pressure remeasurement is $9.421_{-0.75}^{+0.68}~\microbar$, similar to the result obtained by \citet{Arimatsu2020}, ${9.56}_{-0.34}^{+0.52}~\microbar$.

\end{appendix}

\end{document}